\documentclass[aps,pre,amsmath,twocolumn,10pt,superscriptaddress]{revtex4-1}
\usepackage[dvipdfmx]{graphicx}
\usepackage{color}
\usepackage{bm}
\usepackage{braket}
\usepackage{amsmath,amssymb}
\usepackage{mathtools}

\usepackage{physics}
\usepackage[normalem]{ulem}

\begin{document}
\title{Globally-stable and metastable crystal structure enumeration using polynomial machine learning potentials in elemental As, Bi, Ga, In, La, P, Sb, Sn, and Te}
\author{Atsuto \surname{Seko}}
\email{seko@cms.mtl.kyoto-u.ac.jp}
\affiliation{Department of Materials Science and Engineering, Kyoto University, Kyoto 606-8501, Japan}

\date{\today}

\begin{abstract}
Machine learning potentials (MLPs) have become indispensable for conducting accurate large-scale atomistic simulations and for the efficient prediction of crystal structures. 
Polynomial MLPs, defined by polynomial rotational invariants, have been systematically developed for a wide range of elemental, alloy, and ionic systems.
This study introduces a highly efficient and robust methodology for enumerating globally stable and metastable structures utilizing polynomial MLPs.
To develop MLPs that are sufficiently robust for structure enumeration, an iterative process involving random structure searches and subsequent updates to the MLPs is employed. 
This methodology has been systematically applied to elemental systems such as As, Bi, Ga, In, La, P, Sb, Sn, and Te, where numerous local minima present significant competition with the global minimum in terms of energy. 
The proposed approach facilitates robust global structure searches and structure enumerations by markedly accelerating the search process and expanding the search space.
\end{abstract}

\maketitle

\section{Introduction}

Machine learning potentials (MLPs) are becoming increasingly popular for performing accurate calculations that are prohibitively expensive if conducted solely using the density functional theory (DFT) calculation, such as large-scale atomistic simulations
\cite{
Lorenz2004210,
behler2007generalized,
behler2011atom,
han2017deep,
258c531ae5de4f5699e2eec2de51c84f,
PhysRevB.96.014112,
bartok2010gaussian,
PhysRevB.90.104108,
PhysRevX.8.041048,
PhysRevLett.114.096405,
PhysRevB.95.214302,
PhysRevB.90.024101,
PhysRevB.92.054113,
PhysRevMaterials.1.063801,
Thompson2015316,
wood2018extending,
PhysRevMaterials.1.043603,
doi-10.1137-15M1054183,
doi:10.1063/1.5126336,
khorshidi2016amp,
doi-10.1063-1.4930541,
PhysRevB.92.045131,
QUA:QUA24836,
Freitas2022,
PhysRevB.99.014104}.
MLPs represent the short-range part of interatomic interactions with systematic structural features using machine learning models, including artificial neural network models, Gaussian process models, and linear models.
These MLP models are typically estimated using extensive datasets generated by DFT calculations. 
MLPs are designed to be more accurate than conventional interatomic potentials while also being computationally much more efficient than DFT calculations.

Global crystal structure optimizations employing MLPs have become increasingly common (e.g., \cite{PhysRevLett.120.156001,PhysRevB.99.064114,GUBAEV2019148,Kharabadze2022,wakai2023efficient,D3CP02817H}).
In such optimizations, heuristic methods such as multi-start approaches and evolutionary algorithms are typically utilized
\cite{
oganov2019structure,
Pickard_2011,
WANG20122063,
GLASS2006713}. 
To reliably identify global minimum structures using heuristic methods, it is essential to enumerate local minima until no new minima are discovered.
This necessity is highlighted by the fact that stopping criteria in multi-start methods often rely on the number of trials and the count of local minima identified \cite{boender1987bayesian}. 
Furthermore, enumerating local minima with energy values close to the global minimum is crucial for predicting stable structures at finite temperatures and for constructing temperature-dependent phase diagrams. 
MLPs can significantly improve the efficiency of the energy calculations required for such local minimum enumeration.
To robustly enumerate local minima, these MLPs must provide both high predictive accuracy across a broad range of structures and computational efficiency.

This study presents an iterative procedure that uses MLPs and random structure searches to enumerate both global and local minimum structures.
The performance of the current procedure for enumerating global and local minimum structures is systematically evaluated in the elemental systems of As, Bi, Ga, In, La, P, Sb, Sn, and Te, which possess complex potential energy surfaces with numerous local minimum structures that have energy values comparable to the global minimum energy.
The systematic application of this procedure to these complex systems highlights the reliability of the current procedure.

This study employs polynomial MLPs, which have been demonstrated to accurately predict properties across a diverse range of structures in various elemental and alloy systems \cite{PhysRevB.99.214108,PhysRevB.102.174104,doi:10.1063/5.0129045}. 
The polynomial MLPs used in this study are derived from datasets encompassing various prototype structures, their derivatives, and local minima iteratively discovered through random structure searches, as described below.
These polynomial MLPs are expressed as polynomials of rotational invariants, systematically derived from order parameters involving radial and spherical harmonic functions. 
Although the descriptive capacity for potential energy is limited by the use of simple polynomial functions rather than more complex models such as artificial neural networks or Gaussian process models, efficient model estimation is still achievable through linear regression methods, facilitated by powerful linear algebra libraries.
Additionally, force and stress tensor components from DFT training datasets can be incorporated straightforwardly. 
Even when these force and stress tensor components are included as training data, model coefficients can be efficiently estimated using rapid linear regression techniques.

This paper is organized as follows: 
Section \ref{mlp-go:Section-method-MLP} demonstrates the formulation of the polynomial MLP and the current procedure for developing the MLP. 
Section \ref{mlp-go:Section-method-go} introduces the current iterative procedure for performing structure enumeration, which involves the revision of MLPs.
In Sec. \ref{mlp-go:Section-structural-similarity}, a structural similarity used for eliminating duplicate structures in the structure search, which is derived from the formulation of the polynomial MLP, is shown.
Section \ref{mlp-go:Section-results-mlp} provides the prediction errors associated with the polynomial MLPs developed using the current iterative procedure.
Section \ref{mlp-go:Section-results-structures} lists the global and local minimum structures found in the elemental systems and analyzes their preference.
Finally, Section \ref{mlp-go:Section-conclusion} provides a summary of this study.

\section{Polynomial Machine Learning Potentials}
\label{mlp-go:Section-method-MLP}

\subsection{Formulation}
This section provides the formulation of the polynomial MLP in elemental systems, which can be simplified from the formulation in multi-component systems \cite{PhysRevB.102.174104,doi:10.1063/5.0129045}.
The short-range part of the potential energy for a structure, $E$, is assumed to be decomposed as $E = \sum_i E^{(i)}$, where $E^{(i)}$ denotes the contribution of interactions between atom $i$ and its neighboring atoms within a given cutoff radius $r_c$, referred to as the atomic energy.
The atomic energy is then approximately given by a function of invariants $\{d_{m}^{(i)}\}$ with any rotations centered at the position of atom $i$ as
\begin{equation}
\label{mlp-go:Eqn-atomic-energy-features}
E^{(i)} = F \left( d_1^{(i)}, d_2^{(i)}, \cdots \right),
\end{equation}
where $d_{m}^{(i)}$ can be referred to as a structural feature for modeling the potential energy.

The polynomial MLP adopts polynomial invariants of the order parameters representing the neighboring atomic density as structural features and employs polynomial functions as function $F$. 
In other words, the atomic energy is modeled as a polynomial function of polynomial invariants in the polynomial MLP.
Thus, the polynomial MLPs can be regarded as a generalization of embedded atom method (EAM) potentials, modified EAM potentials \cite{PhysRevMaterials.1.063801}, a spectral neighbor analysis potential (SNAP) \cite{Thompson2015316}, and a quadratic SNAP \cite{wood2018extending}.
In addition, linear polynomial models using polynomial invariants are analogous to the formulation of the atomic cluster expansion \cite{PhysRevB.99.014104}.

When the neighboring atomic density is expanded in radial functions $\{f_n\}$ and spherical harmonics $\{Y_{lm}\}$, a $p$th-order polynomial invariant for radial index $n$ and set of angular numbers $\{l_1,l_2,\cdots,l_p\}$ is given by a linear combination of products of $p$ order parameters, expressed as
\begin{widetext}
\begin{equation}
\label{mlp-go:Eqn-invariant-form}
d_{nl_1l_2\cdots l_p,(\sigma)}^{(i)} =
\sum_{m_1,m_2,\cdots, m_p} c^{l_1l_2\cdots l_p,(\sigma)}_{m_1m_2\cdots m_p}
a_{nl_1m_1}^{(i)} a_{nl_2m_2}^{(i)} \cdots a_{nl_pm_p}^{(i)},
\end{equation}
\end{widetext}
where order parameter $a^{(i)}_{nlm}$ is component $nlm$ of the neighboring atomic density of atom $i$.
Coefficient set $\{c^{l_1l_2\cdots l_p,(\sigma)}_{m_1m_2\cdots m_p}\}$ ensures that the linear combinations are invariant for arbitrary rotations, which can be enumerated using group theoretical approaches such as the projection operator method \cite{el-batanouny_wooten_2008,PhysRevB.99.214108}.
In terms of fourth- and higher-order polynomial invariants, multiple linear combinations are linearly independent for most of the set $\{l_1,l_2,\cdots,l_p\}$.
They are distinguished by index $\sigma$ if necessary.

Here, the radial functions are Gaussian-type ones expressed by
\begin{equation}
f_{n}(r)=\exp\left[-\beta_n(r-r_n)^{2}\right] f_c(r),
\end{equation}
where $\beta_n$ and $r_n$ denote given parameters.
Cutoff function $f_c$ ensures the smooth decay of the radial function. 
The current MLP employs a cosine-based cutoff function expressed as
\begin{eqnarray}
f_c(r) = \left\{
\begin{aligned}
& \frac{1}{2} \left[ \cos \left( \pi \frac{r}{r_c} \right) + 1\right] & (r \le r_c)\\
& 0 & (r > r_c)
\end{aligned}
\right ..
\end{eqnarray}
The order parameter of atom $i$, $a_{nlm}^{(i)}$, is approximately evaluated from the neighboring atomic distribution of atom $i$ as
\begin{equation}
a_{nlm}^{(i)} = \sum_{\{j | r_{ij} \leq r_c\} }
f_n(r_{ij}) Y_{lm}^* (\theta_{ij}, \phi_{ij}),
\end{equation}
where $(r_{ij}, \theta_{ij}, \phi_{ij})$ denotes the spherical coordinates of neighboring atom $j$ centered at the position of atom $i$.
Note that this approximation for the order parameters ignores the non-orthonormality of the Gaussian-type radial functions, but it is acceptable in developing the polynomial MLP \cite{PhysRevB.99.214108}.

Given a set of structural features $D^{(i)} = \{d_1^{(i)},d_2^{(i)},\cdots\}$, polynomial function $F_\xi$ composed of all combinations of $\xi$ structural features is represented as
\begin{eqnarray}
\label{mlp-go:Eqn-polynomials}
F_1 \left(D^{(i)}\right) &=& \sum_{s} w_{s} d_{s}^{(i)}, \nonumber \\
F_2 \left(D^{(i)}\right) &=& \sum_{\{st\}} w_{st} d_{s}^{(i)} d_{t}^{(i)}, \\
F_3 \left(D^{(i)}\right) &=& \sum_{\{stu\}} w_{stu} d_{s}^{(i)} d_{t}^{(i)} d_{u}^{(i)} \nonumber,
\end{eqnarray}
where $w$ denotes a regression coefficient.
A polynomial of the polynomial invariants $D^{(i)}$ is then described as
\begin{equation}
\label{Eqn-polynomial-model1}
E^{(i)} = F_1 \left( D^{(i)} \right) + F_2 \left( D^{(i)} \right)
+ F_3 \left( D^{(i)} \right) + \cdots.
\end{equation}
The current model has no constant term, which means that the atomic energy is measured from the energies of isolated atoms.
In addition to the model given by Eqn. (\ref{Eqn-polynomial-model1}), simpler models composed of a linear polynomial of structural features and a polynomial of a subset of the structural features are also introduced, such as
\begin{eqnarray}
\label{Eqn-polynomial-model2}
E^{(i)} &=& F_1 \left( D^{(i)} \right)
+ F_2 \left( D_{\rm pair}^{(i)} \cup D_2^{(i)} \right), 
\end{eqnarray}
where subsets of $D^{(i)}$ are denoted by
\begin{eqnarray}
D_{\rm pair}^{(i)} = \{d_{n0}^{(i)}\}, D_2^{(i)} = \{d_{nll}^{(i)}\}.
\end{eqnarray}

\subsection{Datasets}

\begin{figure}[tbp]
\includegraphics[clip,width=\linewidth]{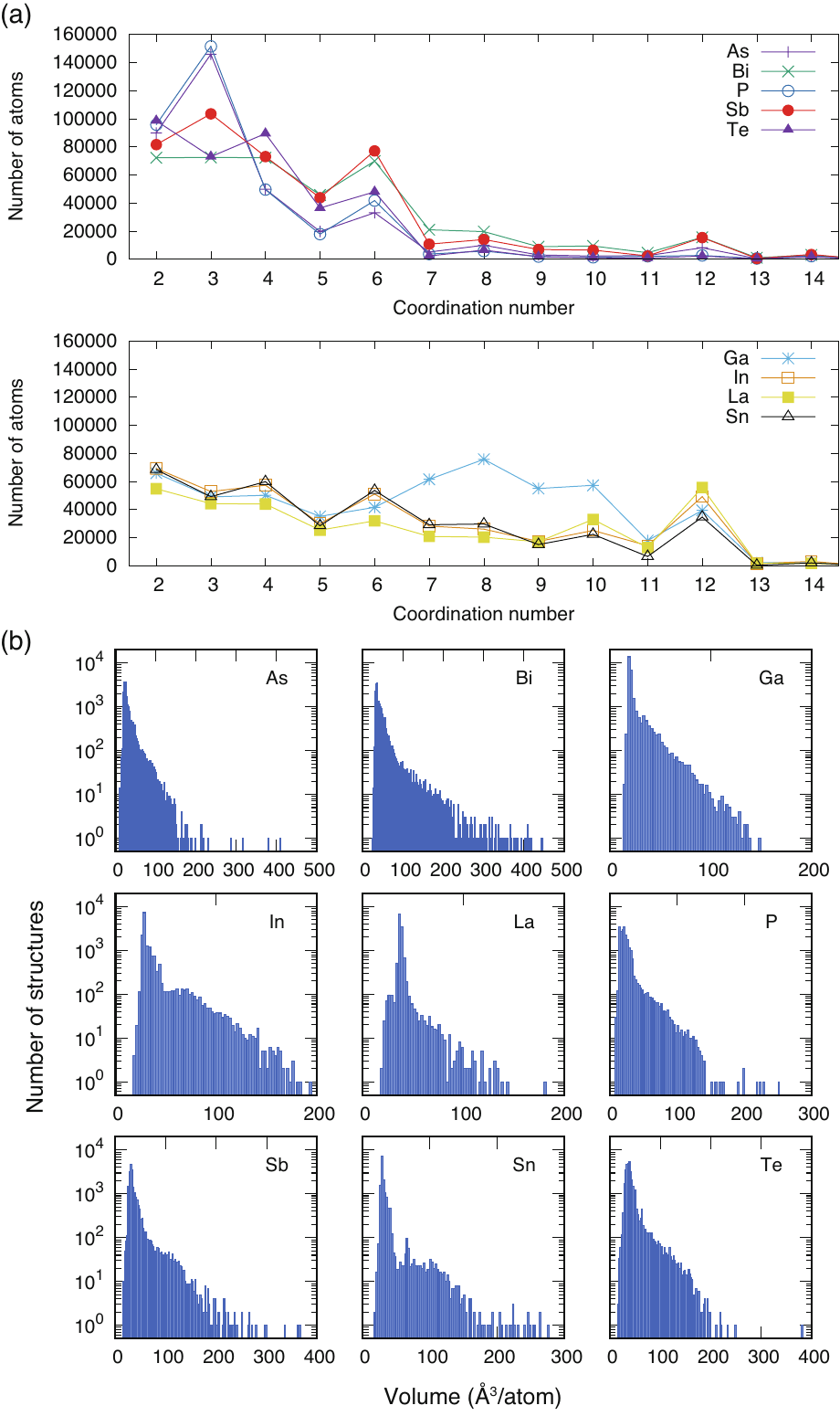}
\caption{
(a) Distribution of the coordination numbers around atoms in structures consisting of the training and test datasets.
(b) Distribution of the volumes for structures in the datasets.
The distribution is displayed in the logarithmic scale for visibility.
}
\label{mlp-go:Fig-dataset}
\end{figure}

Datasets for each elemental system were generated as follows. 
First, the atomic positions and lattice constants of 86 prototype structures \cite{PhysRevB.99.214108} using the DFT calculation were fully optimized. 
They comprise single elements with zero oxidation state from the Inorganic Crystal Structure Database (ICSD) \cite{bergerhoff1987crystal}, including metallic closed-packed structures, covalent structures, layered structures, and structures reported as high-pressure phases. 
Then, 13000--15000 structures were generated by randomly introducing lattice expansions, lattice distortions, and atomic displacements into supercells of optimized prototype structures. 
The entire set of structures was randomly divided into training and test datasets at a ratio of nine to one. 
These datasets were used to develop initial MLPs for the iterative procedure aimed at creating polynomial MLPs robust for structure enumeration, as described in Sec \ref{mlp-go:Sec-go-procedure}. 
During the iterative process, structures predicted to be local minima are incorporated into these datasets.

Figure \ref{mlp-go:Fig-dataset} (a) displays the distribution of coordination numbers around atoms in structures that are present in both the training and test datasets.
The datasets for the elements Ga, In, La, and Sn include structures with coordination numbers ranging from 2 to 12. 
In contrast, the datasets for the elements As, Bi, P, Sn, and Te are biased and predominantly composed of structures with coordination numbers between 2 and 8. 
The distribution of coordination numbers is system-dependent, reflecting the diversity of prototype structures optimized using DFT calculations. 
Even though local geometry optimizations are performed starting from the same prototype structure in different systems, the converged structures may differ due to the varying preferences of the local environment.

Figure \ref{mlp-go:Fig-dataset} (b) shows the volume distribution of the datasets, displayed in a logarithmic scale for better visibility. 
Most of the structures are distributed around the equilibrium volumes of the prototype structures. 
On the other hand, the maximum value of the volume in the distribution is nearly ten times the equilibrium volumes.

DFT calculations were performed for structures in the datasets using the plane-wave-basis projector augmented wave (PAW) method \cite{PAW1} within the Perdew--Burke--Ernzerhof exchange-correlation functional \cite{GGA:PBE96} as implemented in the \textsc{vasp} code \cite{VASP1,VASP2,PAW2}.
The cutoff energy was set to 300 eV.
The allowed spacing between $k$-points was approximately set to 0.09 \AA$^{-1}$.
The total energies converged to less than 10$^{-3}$ meV/supercell.
The atomic positions and lattice constants of the prototype structures were optimized until the residual forces were less than 10$^{-2}$ eV/\AA.
The configurations of the valence electrons in the PAW potentials are 
$4s^2 4p^3$ for As, 
$6s^2 6p^3$ for Bi, 
$4s^2 4p^1$ for Ga, 
$5s^2 5p^1$ for In, 
$5s^2 5p^6 5d^1 6s^2$ for La, 
$3s^2 3p^3$ for P, 
$5s^2 5p^3$ for Sb, 
$5s^2 5p^2$ for Sn, 
and $5s^2 5p^4$ for Te.
These PAW potentials include scalar-relativistic corrections, and spin-orbit coupling was not considered in all the systems.

\subsection{Regression}

Weighted linear ridge regression was used to determine the regression coefficients for potential energy models.
The energy values and force components in the training dataset were employed as observation entries in the regression.
The forces acting on atoms are derived as linear models with the coefficients of the potential energy model, and the predictor matrix $\bm{X}$ and observation vector $\bm{y}$ can be written in a submatrix form as
\begin{equation}
\bm{X} =
\begin{bmatrix}
\bm{X}_{\rm energy} \\
\bm{X}_{\rm force} \\
\end{bmatrix}
,\qquad \bm{y} =
\begin{bmatrix}
\bm{e} \\
\bm{f} \\
\end{bmatrix}.
\end{equation}
The matrix $\bm{X}_{\rm energy}$ comprises structural features and their products, and the elements of $\bm{X}_{\rm force}$ are related to the derivatives of structural features with respect to the Cartesian coordinates of atoms, which were derived in Ref. \onlinecite{PhysRevB.99.214108}.
The observation vector $\bm{y}$ includes components of $\bm{e}$ and $\bm{f}$, which contain the total energies and the forces acting on atoms in the training dataset, respectively. 
These values were obtained from DFT calculations.

Weighted linear ridge regression is a technique that can be used to shrink the regression coefficients by imposing a penalty.
The approach involves minimizing the penalized residual sum of squares, which is given by
\begin{equation}
L(\bm{w}) = ||\bm{W} (\bm{X}\bm{w} - \bm{y}) ||_2^2 + \lambda ||\bm{w}||_2^2,
\end{equation}
where $\lambda$ and $\bm{W}$ denote the magnitude of the penalty and the diagonal matrix where non-zero elements correspond to weights for data entries, respectively.
The solution to the minimization problem is represented by
\begin{equation}
\bm{\hat w} = (\bm{X}^\top \bm{W}^2 \bm{X} + \lambda \bm{I})^{-1} \bm{X}^\top \bm{W}^2 \bm{y},
\end{equation}
where $\bm{I}$ denotes the unit matrix.
The solution can be easily obtained using fast linear algebra algorithms while avoiding the well-known multicollinearity problem.
In this study, the magnitude of the penalty was determined so that the estimated regression coefficients result in the lowest root mean square (RMS) error for an extensive test dataset.

The energy and forces acting on atoms have different units that resemble weights for data entries. 
Specifically, the units eV/cell and eV/\AA\ were used for defining energy and force, respectively.
In addition, weights were applied to data entries depending on their values, which aims to develop robust MLPs for essential structures.
This is achieved by decreasing the influence of less significant data entries that have large values.
The weight for the energy entry of $i$-th structure $W(e_{[i]})$ was given as
\begin{equation}
W(e_{[i]}) =
\begin{cases}
0.1 & (e_{[i]} \geq 0)\\
1 & (e_{[i]} < 0)
\end{cases},
\end{equation}
where $e_{[i]}$ denotes the total energy of structure $i$.
Small weights were applied to energy entries of structures that are less stable than isolated atoms.
The weight for the force entry on $\alpha$-component of atom $j$ in structure $i$, $W(f_{[i],j\alpha})$, was given as
\begin{equation}
W(f_{[i],j\alpha}) =
\begin{dcases}
\frac{\varepsilon}{\left|f_{[i],j\alpha}\right|} & (|f_{[i],j\alpha}| \geq \varepsilon)\\
1 & (|f_{[i],j\alpha}| < \varepsilon)
\end{dcases}.
\end{equation}
Here, $\varepsilon$ was set to 1 eV/\AA.
Smaller weights were assigned to strong force components, whereas larger weights were given to weak force components observed in structures near local minima. 
This weighting strategy can naturally enhance the performance of MLPs in predicting structures close to local minima while reducing the risk of anomalous predictions for structures located far from local minima with high energy values.

\subsection{Model selection}

The accuracy and computational efficiency of MLPs are influenced by several input parameters. 
To determine optimal MLPs that achieve a balance between these factors, a systematic grid search is performed. 
The parameters considered in the grid search include the cutoff radius, the type of potential energy model, the number of radial functions, and the truncation of polynomial invariants, specifically the maximum angular momentum of spherical harmonics $\{l_{\rm max}^{(2)}, l_{\rm max}^{(3)}, l_{\rm max}^{(4)}, \cdots \}$, and the polynomial order of the invariants. 
Since the accuracy and computational efficiency are conflicting properties in polynomial MLPs, Pareto-optimal MLPs are considered suitable candidates for enumerating crystal structures.

For each system, an efficient MLP is selected from the Pareto-optimal candidates. 
The chosen MLP is desired to meet the following several criteria to effectively perform crystal structure enumerations, although it is nearly impossible to develop MLPs that perfectly satisfy all these criteria.
First, the MLP must accurately reconstruct the shape of the potential energy surface, including the locations of local minima, and should be capable of predicting a broad range of local minima structures, ideally encompassing all relevant minima. 
Second, the MLP should be able to identify unrealistic and hypothetical structures with high DFT energy values as energetically unstable. 
Numerous such structures are evaluated during crystal structure enumerations. 
Third, the potential energy surface of the MLP should contain no spurious local minima to ensure efficient structure predictions. 
Lastly, the MLP should perform energy and force calculations efficiently; thus, MLPs requiring approximately 1 ms/atom for a single point calculation are preferred.

\section{Structure enumeration}
\label{mlp-go:Section-method-go}

\subsection{General statement}
In a structure optimization problem with $N$ atoms at a given composition, a crystal structure can be represented by $3N + 3$ independent variables $\bm{x}$. 
They do not contain the translational and rotational degrees of freedom. 
The goal of global structure optimization is to minimize the structure-dependent potential energy $E(\bm{x})$, formulated as 
\begin{equation}
E^* = \min_{\bm{x} \in \mathbb{D}} E(\bm{x}),
\end{equation}
within the nonempty feasible region $\mathbb{D}$ for crystal structure representation $\bm{x}$.
The crystal structure that yields the minimum energy $E^*$ is denoted as the global minimum structure $\bm{x}^*$, hence $E^* = E(\bm{x}^*)$. 
Similarly, a local minimum structure is denoted as $\bm{x}_l^*$, and a set of local minimum structures with energy values lower than a given energy threshold $\theta$ is denoted as
\begin{equation}
\mathbb{D}^*_\theta = \left \{ \bm{x}_l^* \in \mathbb{D} \:|\: E(\bm{x}_l^*) \leq \theta \right \}.
\end{equation}

Depending on the choice of the crystal structure representation, multiple representations can correspond to the same structure, making the entire feasible region redundant. 
In addition, there may be subregions within the feasible region where it is impossible to form crystal structures.
Furthermore, the feasible region contains physically unrealistic structures. 
To search for crystal structures efficiently, these unrealistic structures are often eliminated from the feasible region in advance.
These subregions can be defined by constraints applied to the entire feasible region, particularly linear constraints.
Given a set of constraints $\{g_1(\bm{x}) \leq 0, \cdots, g_m(\bm{x}) \leq 0\}$, the reduced feasible region $\Delta$ is described as
\begin{equation}
\label{mlp-go:Eq-reduced-region}
\Delta = \bigcap_m \Delta_m, 
\end{equation}
where $\Delta_m$ denotes the feasible region reduced by constraint $g_m(\bm{x}) \leq 0$, written as
\begin{equation}
\Delta_m = \{\bm{x} \in \mathbb{D} \:| \: g_m(\bm{x}) \leq 0 \}.
\end{equation}
When the feasible region is reduced, the problems of finding the global minimum structure and enumerating local minimum structures are restated as
\begin{equation}
E^* = \min_{\bm{x} \in \Delta} E(\bm{x})
\end{equation}
and
\begin{equation}
\Delta^*_\theta = \left \{ \bm{x}_l^* \in \Delta \:|\: E(\bm{x}_l^*) \leq \theta \right \},
\end{equation}
respectively.

\subsection{Computational procedures}
\label{mlp-go:Sec-go-procedure}

This study employs a random structure search method \cite{Pickard_2011}, which corresponds to the multi-start approach for global optimization \cite{hendrix2010introduction}. 
This method involves repeatedly performing local geometry optimizations starting from uniformly sampled structures within the feasible region $\mathbb{D}$ or a reduced feasible region $\Delta$. 
The random structure search is an effective method for enumerating both global and local minimum structures, owing to its uniform sampling strategy, straightforward implementation, and ease of parallelization.

In the current procedure, most of the $ab$ $initio$ calculations in the $ab$ $initio$ random structure search (AIRSS) method \cite{Pickard_2011} are replaced with MLP calculations.
However, there are concerns associated with using MLPs for crystal structure enumeration.
Firstly, MLPs often exhibit significant prediction errors for structures that lie outside the training dataset and sometimes produce spurious local minimum structures.
Since it is infeasible to prepare a training dataset that covers all possible local minimum structures in advance, an iterative approach involving updates to polynomial MLPs is employed. 
Secondly, even with iterative updates to polynomial MLPs that reduce prediction errors, some small but non-negligible errors may persist. 
This can complicate the assessment of stability between local minimum structures based on energy values predicted by MLPs.
Consequently, the final assessment of stability between local minimum structures is carried out using DFT calculations.
Local geometry optimizations for the local minimum structures identified by the MLP are systematically performed using DFT calculations.

The following is the current procedure for finding the global and local minimum structures: 
(1) A polynomial MLP is developed for the chosen model using DFT training and test datasets. 
(2) A large number of initial structures ($\sim 10^5$) are randomly and uniformly sampled from the reduced feasible region $\Delta$, which is limited to structures with up to twelve atoms. 
(3) Local geometry optimizations are systematically performed on the initial structures using the polynomial MLP. 
(4) Duplicate local minimum structures are identified and removed using space-group identification and a similarity measure that is closely related to the polynomial MLP (as described in Sec. \ref{mlp-go:Section-structural-similarity}).
A small tolerance value is used to eliminate duplicate structures, while ensuring that all borderline structures are kept as candidates for local minimum structures.
(5) Single-point DFT calculations are performed for local minimum structures with low energy values, and their results are added to the training dataset.
Steps (1)--(5) are repeated iteratively until a robust random structure search becomes possible.
(6) Local geometry optimizations are then performed using the DFT calculation. 
These optimizations begin with a subset of local minimum structures with MLP energy values lower than a given threshold value $\theta^{\rm MLP}$.
(7) Duplicate local minimum structures are again identified and removed using a tolerance value that is slightly larger than the one used in step (4).

The feasible region is defined by the metric tensor of lattice basis vectors and the fractional coordinates of the atomic positions with respect to the lattice basis vectors.
This feasible region is then reduced by applying the main conditions for defining the Niggli reduced cell \cite{ITA2002}.
To create initial structures, a set of three diagonal components of the metric tensor is uniformly sampled, given an upper bound for the diagonal elements.
Following the main conditions for the Niggli cell determined by the diagonal components, the non-diagonal components of the metric tensor are then randomly sampled between the upper and lower bounds.
Although the fractional coordinates of the atomic positions in the reduced feasible region have duplicate subregions, the efficiency of the random structure search remains unaffected.

Note that the AIRSS can be computationally expensive because it requires numerous local geometry optimizations using $ab$ $initio$ calculations. 
To reduce the computational cost, initial structures are usually sampled within a feasible region based on prior knowledge of crystal structure stability. 
On the other hand, the current approach does not rely on any prior knowledge about the initial crystal structure, except for the upper bound of the metric tensor of lattice basis vectors. 
This approach allows for robust enumeration of both global and local minimum structures without preconceived assumptions about the initial structures.

\subsection{Duplicate structure elimination}
\label{mlp-go:Section-structural-similarity}

Numerous local geometry optimizations, as outlined in steps (3) and (6) of the current procedure described in Sec. \ref{mlp-go:Sec-go-procedure}, can lead to the generation of many duplicate structures. 
This issue has also been noted in global structure searches conducted in many other studies.
In this study, both space group identification and a distance measure based on structural features are employed to eliminate duplicate structures. 
The structural features, detailed in Eq. (\ref{mlp-go:Eqn-invariant-form}), are derived from the selected polynomial MLP. 
These features are then normalized using the regression coefficients of the chosen polynomial MLP.

When indices of the structural features are written by single index $s = \{n,l_1,l_2,\cdots,l_p,(\sigma)\}$, the averages of the normalized structural features and their products over atoms in structure $i$, $\bm{\bar d}_{[i]}$, are described as
\begin{equation}
\bm{\bar d}_{[i]} = \left[\cdots,\bar d_{[i],s},\cdots,\bar d_{[i],st},\cdots,\bar d_{[i],stu},\cdots\right],
\end{equation}
where
\begin{eqnarray}
\bar d_{[i],s} &=& \frac{1}{N_{[i]}} \sum_k w_s d_{[i],s}^{(k)}, \nonumber \\
\bar d_{[i],st} &=& \frac{1}{N_{[i]}} \sum_k w_{st} d_{[i],s}^{(k)} d_{[i],t}^{(k)}, \\
\bar d_{[i],stu} &=& \frac{1}{N_{[i]}} \sum_k w_{stu} d_{[i],s}^{(k)} d_{[i],t}^{(k)} d_{[i],u}^{(k)},\nonumber
\end{eqnarray}
and $N_{[i]}$ denotes the number of atoms in structure $i$.
The distance between structures, $i$ and $j$, is then measured as the Euclidean distance of their corresponding $\bm{\bar d}_{[i]}$ and $\bm{\bar d}_{[j]}$, $\|\bm{\bar d}_{[i]}-\bm{\bar d}_{[j]}\|_2$.
If the distance between the structures is small, they are considered similar in terms of the local neighboring atomic distribution.
It is important to note that the current set of structural features is a generalization of radial distribution functions and angular distribution functions. 
These functions represent local neighboring atomic distributions and have often been used to eliminate duplicate structures (e.g., \cite{valle2010crystal}).

In the current structure searches using the MLPs, the number of local minimum structures frequently exceeds 50,000, which makes it impractical to compute distances for all pairs of structures. 
Therefore, an efficient procedure is introduced to eliminate duplicate structures.
(1) Structures with different energy values are considered to be different.
The energy value can act as a concise key to determine whether structures are different or not. 
This is based on the fact that the energy of structure $i$ is expressed using its normalized structural features as
\begin{equation}
E_{[i]} = \sum_s \bar d_{[i],s} + \sum_{st} \bar d_{[i],st} + \sum_{stu} \bar d_{[i],stu} + \cdots.
\end{equation}
If the absolute difference between the energy values of structures $i$ and $j$ is more than a given tolerance parameter $\varepsilon_E$, $|E_{[i]} - E_{[j]}| > \varepsilon_E$, they are regarded as different.
Here, the tolerance parameter is set to $\varepsilon_E = 0.01$ meV/atom.
If $|E_{[i]} - E_{[j]}| \leq \varepsilon_E$, we proceed to step (2).
(2) Structures with different space groups are considered different.
Because the space group identification depends on the tolerance parameter, multiple tolerance parameters are used. 
A single structure is identified with the set of space groups from the multiple tolerance parameters.
If the sets of space groups for structures $i$ and $j$, $\{\mathbb{G}\}_{[i]}$ and $\{\mathbb{G}\}_{[j]}$, have no common elements, i.e., $\{\mathbb{G}\}_{[i]} \cap \{\mathbb{G}\}_{[j]} = \varnothing$, the structures are regarded to be different.
Otherwise, we proceed to step (3).
(3) Two structures separated by a distance larger than a tolerance parameter $\varepsilon_d$, i.e., $\|\bm{\bar d}_{[i]}-\bm{\bar d}_{[j]}\|_2 > \varepsilon_d$, are considered different.
Here, the tolerance parameter is set to $\varepsilon_d = 0.001$.
Otherwise, they are regarded to be identical.

Note that the distances between structures that are clearly identical are less than $10^{-7}$.
In contrast, the distances between structures that are visibly different but have similar energy values are greater than 0.1. 
Most structure pairs fall into either of these categories, although some pairs may lie near the borderline. 
For local minimum structures obtained through DFT calculations, tolerance parameters of
$\varepsilon_E = 1$ meV/atom and $\varepsilon_d = 0.01$ are used. 
These values are larger than those applied to structures obtained using MLPs due to the larger convergence criteria employed in local geometry optimizations with DFT.

\section{Results and discussion}
\subsection{Machine learning potentials}
\label{mlp-go:Section-results-mlp}

\begin{table}[tbp]
\begin{ruledtabular}
\caption{
Accuracy and computational efficiency of the polynomial MLPs used for enumerating the global and local minimum structures.
The RMS errors are calculated for test datasets, excluding structures with exceptionally high positive energy values.
The computational efficiency is evaluated by measuring the elapsed time to compute the energy, forces, and stress tensors of a structure consisting of many atoms.
The elapsed time is normalized by the number of atoms because it is proportional to the number of atoms.
The elapsed time for a single point calculation is estimated using a single core of Intel\textregistered\ Xeon\textregistered\ E5-2695 v4 (2.10 GHz) and an implementation of the polynomial MLP to the \textsc{lammps} code \cite{LammpsPolyMLP}.
}
\label{mlp-go:Table-mlp-error}
\begin{tabular}{cccc}
& RMS error (energy) & RMS error (force) & Time \\
& (meV/atom) & (eV/\AA) & (ms/atom/step) \\
\hline
As  & 5.7 & 0.117 & 1.4 \\ 
Bi  & 4.0 & 0.065 & 1.0 \\ 
Ga  & 2.1 & 0.030 & 0.6 \\ 
In  & 1.7 & 0.019 & 0.5 \\ 
La  & 3.1 & 0.052 & 1.0 \\ 
P   & 6.2 & 0.142 & 1.3 \\ 
Sb  & 4.5 & 0.075 & 1.4 \\ 
Sn  & 2.6 & 0.037 & 1.0 \\ 
Te  & 6.0 & 0.098 & 2.4    
\end{tabular}
\end{ruledtabular}
\end{table}

\begin{figure}[tbp]
\includegraphics[clip,width=\linewidth]{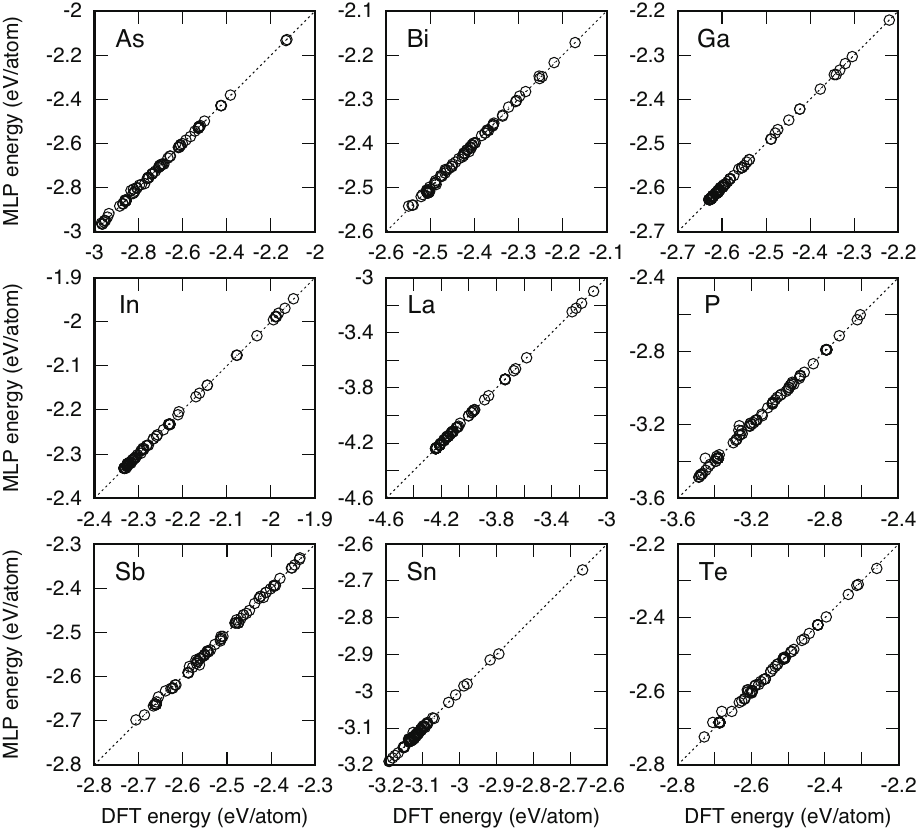}
\caption{
Prediction errors of the selected MLPs for 86 prototype structures.
}
\label{mlp-go:Fig-icsd-predictions}
\end{figure}

\begin{figure}[tbp]
\includegraphics[clip,width=\linewidth]{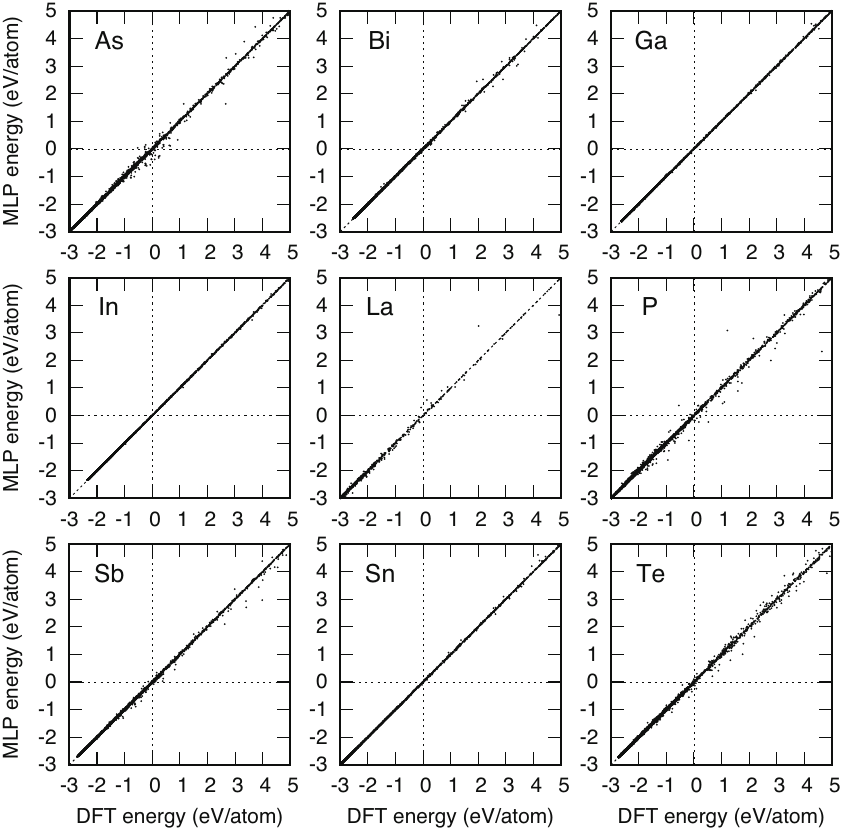}
\caption{
Distribution of the energy values for the training and test datasets, predicted using the DFT calculation and the selected MLPs.
}
\label{mlp-go:Fig-energy-distributions}
\end{figure}

Table \ref{mlp-go:Table-mlp-error} lists the RMS errors of the updated MLPs at the final iteration.
The RMS errors are assessed for the test datasets, excluding structures with exceptionally high positive energy values, which offers a more practical measure of the accuracy.
Figure \ref{mlp-go:Fig-icsd-predictions} shows the prediction errors of the selected MLPs for 86 prototype structures.
As depicted in Fig. \ref{mlp-go:Fig-icsd-predictions}, the MLPs have a high predictive power for various structures.
Figure \ref{mlp-go:Fig-energy-distributions} displays the distributions of energies of structures in both the training and test datasets, computed using the DFT calculation and the polynomial MLP.
The distributions demonstrate that the polynomial MLP is accurate for many typical structures and their derivatives containing diverse neighborhood environments and coordination numbers, as evidenced by a narrow distribution of errors. 
Overall, these results indicate that the polynomial MLPs are reliable for predicting the energies of a wide range of structures.

Polynomial MLPs developed using the final datasets including local minimum structures can be accessed through the \textsc{Polynomial machine learning potential repository} \cite{doi:10.1063/5.0129045,MachineLearningPotentialRepository}. 
While only the predictive power for energy has been demonstrated in this study, the repository website includes predictions for other properties as well. 
Additionally, the predictive power for liquid states in the elemental Bi, Ga, In, and Sn can be found elsewhere \cite{PhysRevB.109.214207}.

\subsection{Structure enumeration}
\label{mlp-go:Section-results-structures}

Structure enumeration begins with polynomial MLPs developed according to the procedure outlined in Sec. \ref{mlp-go:Sec-go-procedure}.
These MLPs are constructed from datasets that include various prototype structures, their derivatives, and structures predicted to be local minima by the MLPs. 
Subsequently, random structure searches are performed using these MLPs, involving systematic repetitions of local geometry optimizations.
From these searches, unique local minimum structures with MLP energy values, $E^{\rm MLP}$, lower than a given energy threshold $\theta^{\rm MLP}$, $E^{\rm MLP} \leq \theta^{\rm MLP}$, are extracted.
To ensure reliable structure enumerations, the following threshold values are used:  25 meV/atom for Ga, 50 meV/atom for As, Bi,  and In, 75 meV/atom for La, P, Sb, and Sn, and 100 meV/atom for Te, which are measured from the lowest energy. 

When it is assumed that the Bayesian estimates on the number of local minima in multi-start approaches \cite{boender1987bayesian} are applicable to the current structure search, the estimates are calculated for $E^{\rm MLP} \leq \theta^{\rm MLP}$ as follows: 465.9, 189.8, 1566.7, 520.1, 298.9, 282.7, 407.6, 403.3, 418.9 for As, Bi, Ga, In, La, P, Sb, Sn, and Te, respectively. 
These estimates are very close to the numbers of the local minimum structures found in the current random structure searches, i.e., 464, 189, 1529, 516, 297, 282, 406, 400, and 417, respectively.
This indicates that most of the local minimum structures with $E^{\rm MLP} \leq \theta^{\rm MLP}$ existing in the feasible region should be found in the current structure searches using the MLPs, including the globally stable structure.

\begin{figure}[tbp]
\includegraphics[clip,width=\linewidth]{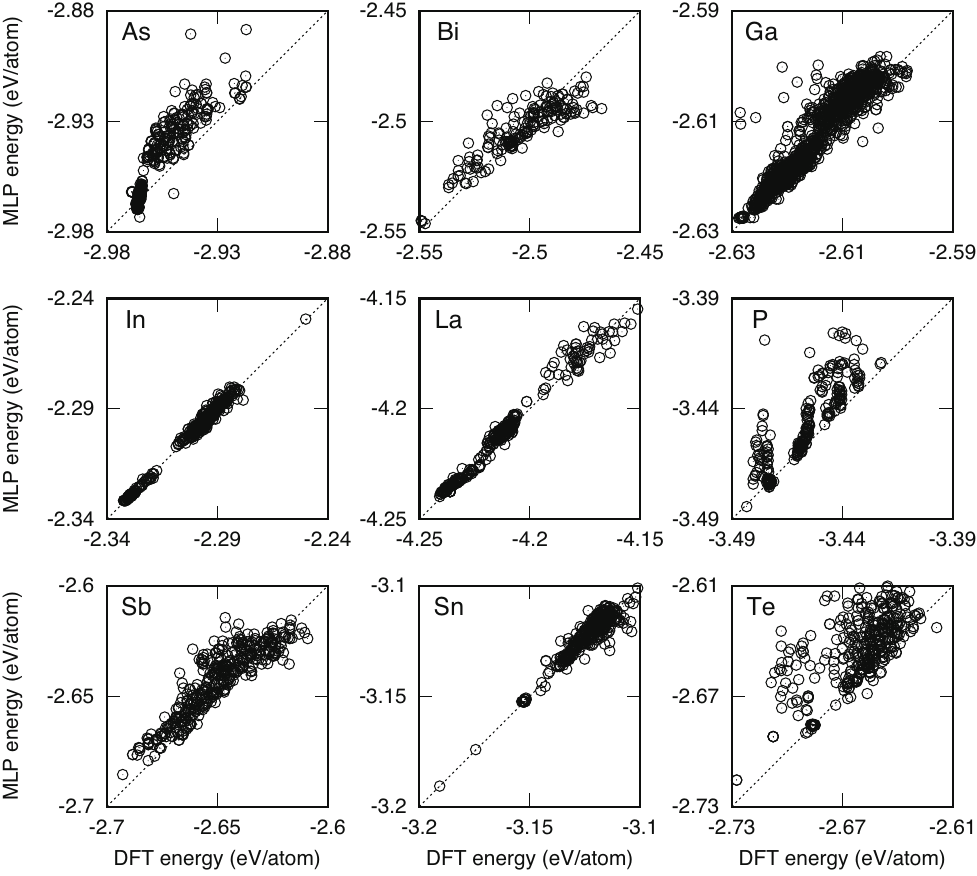}
\caption{
Energy distribution for the global and local minimum structures.
The MLP energy is calculated using the updated MLP at the final iteration.
The energy scales in this figure are significantly smaller compared to those used in Figs. \ref{mlp-go:Fig-icsd-predictions} and \ref{mlp-go:Fig-energy-distributions}.
}
\label{mlp-go:Fig-locals-dft-mlp}
\end{figure}

Once a set of local minimum structures with $E^{\rm MLP} \leq \theta^{\rm MLP}$ is obtained using the MLP, local geometry optimizations are performed using the DFT calculation. 
This allows us to evaluate the stability of the structures more accurately. 
Figure \ref{mlp-go:Fig-locals-dft-mlp} shows the energy distribution of the local minimum structures in the nine elemental systems.
The energy values of the local minimum structures predicted using the DFT calculation and the polynomial MLP show strong correlations as a result of the iterative update of the MLPs.
This result suggests that the updated MLPs can enumerate most of the local minimum structures with low energy values.
However, there are some local minimum structures whose energies predicted by the polynomial MLPs deviate from their DFT energies.
While the MLPs are less reliable in predicting the energy of such structures, they can still play an adequate role in finding local minimum structures with low energy values. 
If we aim to improve the predictive power for such structures, more complex models for the MLPs are necessary.

\begin{table}[tbp]
\begin{ruledtabular}
\caption{
Number of the global and local minimum structures for several energy threshold values $\theta$.
The energy values of the local minimum structures are computed using the DFT calculation.
The energy threshold values are described in the unit of meV/atom.
}
\label{mlp-go:Table-number-locals}
\begin{tabular}{cccccc}
& $\theta=5$ & $\theta=10$ & $\theta=15$ & $\theta=20$ & $\theta=25$ \\
\hline
Ga  & 1 & 101 & 616 & 902 & 1142 \\ 
\hline
& $\theta=10$ & $\theta=20$ & $\theta=30$ & $\theta=40$ & $\theta=50$ \\
\hline
As  & 269 & 336 & 387 & 407 & 415 \\ 
Bi  & 2 & 11 & 27 & 54 & 101 \\ 
In  & 99 & 113 & 135 & 336 & 474 \\ 
\hline
& $\theta=15$ & $\theta=30$ & $\theta=45$ & $\theta=60$ & $\theta=75$ \\
\hline
La  & 73 & 150 & 205 & 220 & 253 \\ 
P   & 112 & 191 & 255 & 274 & 276 \\ 
Sb  & 2 & 24 & 101 & 220 & 302 \\ 
Sn  & 1 & 4 & 13 & 81 & 310 \\ 
\hline
& $\theta=20$ & $\theta=40$ & $\theta=60$ & $\theta=80$ & $\theta=100$ \\
\hline
Te  & 7 & 41 & 81 & 253 & 339  
\end{tabular}
\end{ruledtabular}
\end{table}

Table \ref{mlp-go:Table-number-locals} presents the number of unique local minimum structures with DFT energy values below various thresholds. 
It is observed that many local minimum structures have energy values very close to the global minimum energy in these systems. 
The set of global and local minimum structures identified for a small threshold is reliable; however, the set of global and local minimum structures for thresholds near $\theta^{\rm MLP}$ may be incomplete. 
Due to prediction errors of the MLPs, some structures with MLP energy values exceeding $\theta^{\rm MLP}$ may still exhibit DFT energy values below the thresholds.

Furthermore, the number of local minimum structures identified using DFT calculations is smaller than that obtained using MLPs, even though the majority of structures identified through MLPs converge to unique local minima upon DFT geometry optimization. 
This discrepancy occurs because some structures with MLP energy values below $\theta^{\rm MLP}$ may have DFT energy values exceeding $\theta^{\rm MLP}$. 
Additionally, the reduction in the number of structures is influenced by the tolerance parameters used to eliminate duplicate structures.
Specifically, a larger tolerance value is applied to structures identified through DFT calculations compared to those obtained using MLPs, as detailed in Section \ref{mlp-go:Sec-go-procedure}.

\subsubsection{As}

\begin{figure}[tbp]
\includegraphics[clip,width=\linewidth]{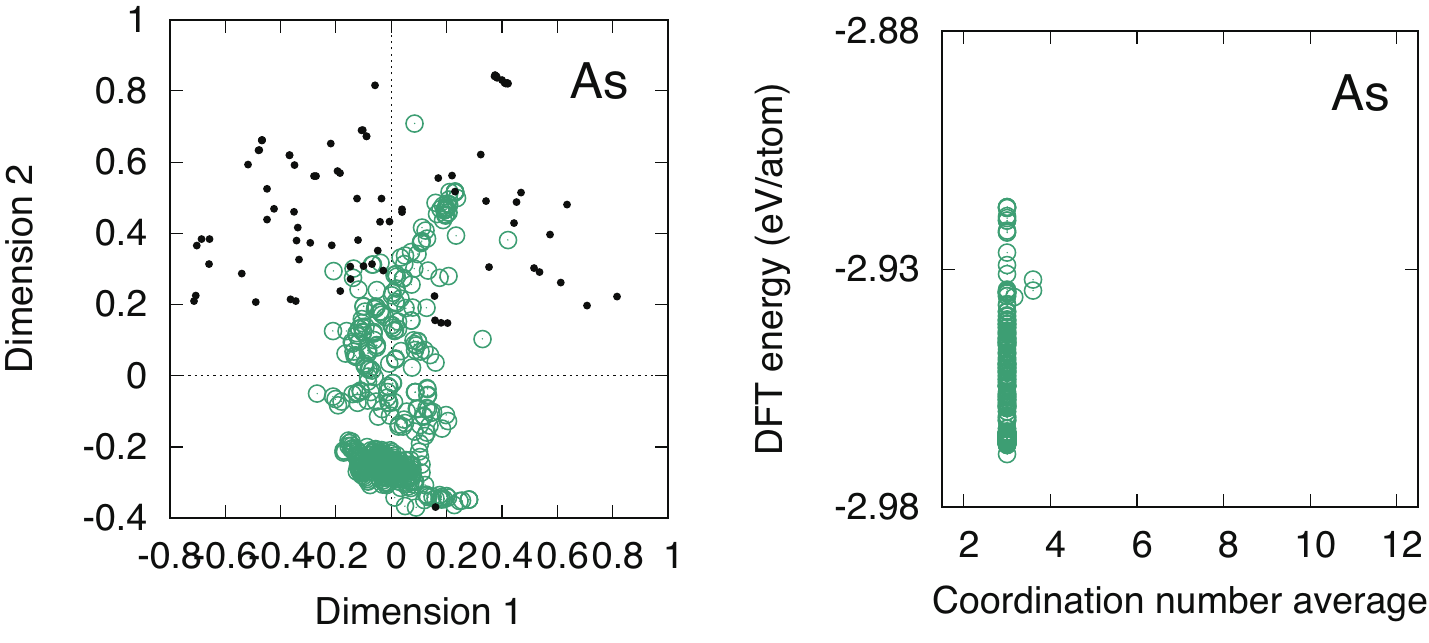}
\caption{
(a) Distribution of the local minimum structures and equilibrium prototype structures in As.
The prototype structures are used to derive structures in the training dataset.
These structures are mapped into a two-dimensional plane using an MDS for visibility.
The open and closed circles show the local minimum structures and prototype structures, respectively.
(b) Average coordination numbers of the local minimum structures.
The cutoff radius is given as 1.2 times the nearest neighbor distance.
}
\label{mlp-go:Fig-structure-As}
\end{figure}

In the elemental As, it is found that there are 415 local minimum structures that have relative energy values lower than $\theta = 50$ meV/atom.
Specifically, 269 of those structures have relative energy values less than $\theta = 10$ meV/atom.
Figure \ref{mlp-go:Fig-structure-As} (a) displays the distribution of local minimum structures and equilibrium prototype structures that are used to create structures in the training dataset.
The structural features of the polynomial MLP for these structures are mapped onto a two-dimensional plane using multi-dimensional scaling (MDS) \cite{borg2005modern}.
Therefore, this figure provides a visual representation of the similarity measure between the local minimum structures and training dataset. 
As shown in the figure, many local minimum structures are located far from the prototype structures.
Figure \ref{mlp-go:Fig-structure-As} (b) shows the distribution of average coordination numbers for the local minimum structures.
The cutoff radius used is 1.2 times the nearest neighbor distance.
The global minimum structure and most of the local minimum structures are three-coordinated, which is consistent with the experimental structure.

\begin{table}[tbp]
\begin{ruledtabular}
\caption{
Global and local minimum structures in As.
The relative energy $\Delta E$ calculated using the DFT calculation is expressed in the unit of meV/atom.
$Z$ denotes the number of atoms included in the unitcell.
Parentheses in the columns of ICSD identifiers and prototype structures indicate that the corresponding structure is close to the prototype in the parentheses.
However, its space group is different from that of the prototype structure and/or large tolerance values are required for the prototype identification.
}
\label{mlp-go:Table-structure-As}
\begin{tabular}{ccccc}
Space group & ICSD-ID & Prototype & $Z$ & $\Delta E$ \\
\hline
$R\bar3m$       &  9859  &  As   & 6   & 0.0 \\ 
$P\bar3m1$      & (1425) & (HCP) & 2   & 1.9 \\ 
$C2/m$          & (1425) & (HCP) & 8   & 2.0 \\ 
$P\bar3m1$      & (1425) & (HCP) & 4   & 2.0 \\ 
$R\bar3m$       &  $-$   &  $-$  & 24  & 2.1 \\ 
$R\bar3m$       &  $-$   &  $-$  & 30  & 2.1 \\ 
$P3m1$          &  $-$   &  $-$  & 6   & 2.1 \\ 
$P3m1$          &(31170) &(C($P6_3mc$))& 4   & 2.1 \\ 
$R\bar3m$       &  $-$   &  $-$  & 18  & 2.2 \\ 
$R\bar3m$       &  $-$   &  $-$  & 12  & 2.2 \\ 
$P6_3mc$        & 31170  &C($P6_3mc$)& 4   & 2.2 \\ 
$P\bar3m1$      &  $-$   &  $-$  & 8   & 2.2 \\ 
$R3m$           &  $-$   &  $-$  & 12  & 2.2 \\ 
$R\bar3m$       &  $-$   &  $-$  & 18  & 2.2 \\ 
$R\bar3m$       &  $-$   &  $-$  & 18  & 2.2 \\ 
$R\bar3m$       &  $-$   &  $-$  & 12  & 2.3 \\ 
$R\bar3m$       &  $-$   &  $-$  & 18  & 2.3 \\ 
$R3m$           &  $-$   &  $-$  & 24  & 2.3 \\ 
$R3m$           &  $-$   &  $-$  & 24  & 2.4 \\ 
$R3m$           &  $-$   &  $-$  & 30  & 2.4 \\ 
$R3m$           &  $-$   &  $-$  & 24  & 2.4 \\ 
$R3m$           &  $-$   &  $-$  & 24  & 2.5 \\ 
$R3m$           &  $-$   &  $-$  & 24  & 2.5 \\ 
$R3m$           &  $-$   &  $-$  & 30  & 2.5 \\ 
$R3m$           &  $-$   &  $-$  & 36  & 2.5 \\ 
$P3m1$          &  $-$   &  $-$  & 10  & 2.5 \\ 
$P1$            &  $-$   &  $-$  & 10  & 2.5 \\ 
$R3m$           &  $-$   &  $-$  & 30  & 2.5 \\ 
$R3m$           &  $-$   &  $-$  & 18  & 2.5 \\ 
\hline
$C2/m$          & 2284   &O$_2$($mS4$)& 4   & 3.2 \\ 
$R\bar3m$       & 29213  &Graphite(3R)& 6   & 3.2 \\ 
$R\bar3m$       & 29213  &Graphite(3R)& 6   & 3.8 \\ 
$P6_3mc$        & 31170  &C($P6_3mc$)& 4   & 4.1 \\ 
$Cmce$          & 23836  &black-P& 8   & 18.7 \\ 
\end{tabular}
\end{ruledtabular}
\end{table}

Table \ref{mlp-go:Table-structure-As} provides a list of local minimum structures found in the elemental As.
This list includes only the structures that have low energy values and can be identified with prototype structures and experimental structures in As.
Prototype identifiers that are consistent with the ICSD are given.
The global minimum structure is the three-coordinated As-type structure, which is consistent with the experimental structure at low temperatures \cite{schiferl1969crystal,young1991phase}.
Apart from the global minimum structure, three-coordinated structures such as C($P6_3mc$)-, graphite(3R)-, and black-P-type are also found as local minimum structures.
Among these structures, only the black-P-type structure is an experimental polymorph \cite{smith1975structures}.

\subsubsection{Bi}

\begin{figure}[tbp]
\includegraphics[clip,width=\linewidth]{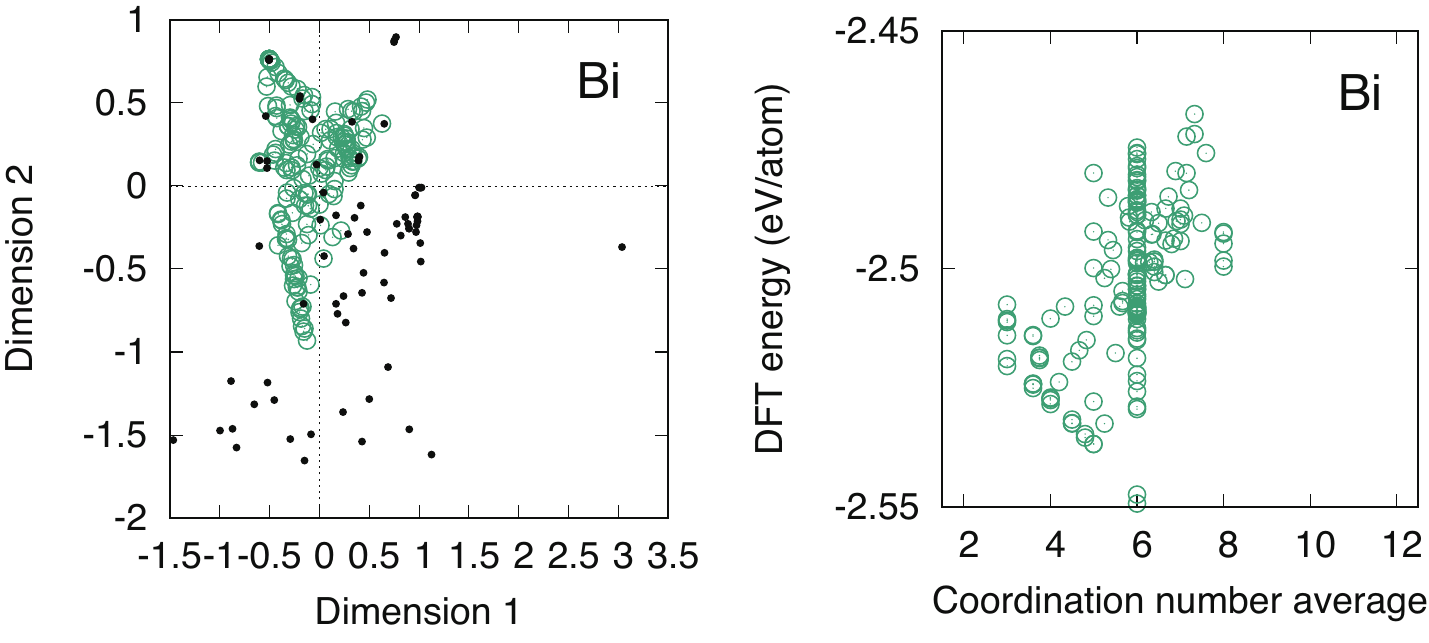}
\caption{
(a) Distribution of the local minimum structures and equilibrium prototype structures in Bi.
These structures are mapped into a two-dimensional plane using an MDS.
The open and closed circles show the local minimum structures and prototype structures, respectively.
(b) Average coordination numbers of the local minimum structures.
The cutoff radius is given as 1.2 times the nearest neighbor distance.
}
\label{mlp-go:Fig-structure-Bi}
\end{figure}

In the elemental Bi, 101 different local minimum structures with relative energy values lower than $\theta =$ 50 meV/atom have been discovered.
The distribution of these local minimum structures and equilibrium prototype structures in Bi is shown in Fig. \ref{mlp-go:Fig-structure-Bi} (a).
The majority of the local minimum structures are located around the equilibrium prototype structures.
Figure \ref{mlp-go:Fig-structure-Bi} (b) shows the distribution of average coordination numbers for the local minimum structures in Bi.
The global minimum structure is six-coordinated, and the average coordination numbers of the local minimum structures in Bi range from three to eight.

\begin{table}[tbp]
\begin{ruledtabular}
\caption{
Global and local minimum structures in Bi.
The DFT relative energy $\Delta E$ is expressed in the unit of meV/atom.
$Z$ denotes the number of atoms included in the unitcell.
}
\label{mlp-go:Table-structure-Bi}
\begin{tabular}{ccccc}
Space group & ICSD-ID & Prototype & $Z$ & $\Delta E$ \\
\hline
$R\bar3m$       &  9859  &  As   & 6   & 0.0 \\ 
$Imma$          & (9859) & (As)  & 4   & 1.8 \\ 
$R\bar3m$       &  $-$   &  $-$  & 36  & 12.5 \\ 
$P\bar3m1$      &  $-$   &  $-$  & 12  & 12.5 \\ 
$P3m1$          &  $-$   &  $-$  & 10  & 13.8 \\ 
$P\bar3m1$      &  $-$   &  $-$  & 10  & 14.6 \\ 
$R3m$           &  $-$   &  $-$  & 24  & 16.7 \\ 
$P\bar3m1$      &  $-$   &  $-$  & 8   & 16.7 \\ 
$P6_3mc$        &  $-$   &  $-$  & 8   & 16.8 \\ 
$R\bar3m$       &  $-$   &  $-$  & 24  & 17.6 \\ 
$C2/m$          &  $-$   &  $-$  & 24  & 19.8 \\ 
$Cm$            &  $-$   &  $-$  & 24  & 20.4 \\ 
$R3m$           &  $-$   &  $-$  & 18  & 20.9 \\ 
$R\bar3m$       &  $-$   &  $-$  & 18  & 21.4 \\ 
$R3m$           &  $-$   &  $-$  & 36  & 21.8 \\ 
$R\bar3m$       &  $-$   &  $-$  & 18  & 22.0 \\ 
$P\bar3m1$      &  $-$   &  $-$  & 6   & 22.2 \\ 
$P2_1/m$        &  $-$   &  $-$  & 10  & 23.5 \\ 
$R3m$           &  $-$   &  $-$  & 30  & 24.2 \\ 
$R3m$           &  $-$   &  $-$  & 30  & 25.0 \\ 
$P3m1$          &  $-$   &  $-$  & 10  & 25.0 \\ 
\hline
$P6_3mc$        & 31180  &C($P6_3mc$) & 4   & 28.9 \\ 
$P2_1/m$        &(43211) & (SC)  & 8   & 38.7 \\ 
$C2$            &(43211) & (SC)  & 18  & 39.6 \\ 
$P1$            &(43211) & (SC)  & 12  & 39.7 \\ 
$Pm\bar3m$      & 43211  &  SC   & 1   & 40.0 \\ 
$I4_1/amd$      & 40037  & $\beta$-Sn & 4   & 41.1 \\ 
$Pnma$          & 165995 &  MnP    & 8   & 41.6 \\ 
$C2/m$          & 409752 & Bi(II)  & 20  & 42.9 \\ 
$Cmce$          & 23836  & black-P & 8   & 49.4 \\ 
$I4_1/amd$      & 109018 & Cs(HP)  & 4   & 49.8 \\ 
\end{tabular}
\end{ruledtabular}
\end{table}

Table \ref{mlp-go:Table-structure-Bi} lists the local minimum structures found in the elemental Bi.
The global minimum structure is the As-type structure, which is consistent with the experimental structure observed at low temperatures \cite{PhysRev.25.753,cucka1962crystal,young1991phase}.
It is worth noting that the global minimum structure can be considered as six-coordinated, which is different from the three-coordinate As-type structure observed in the elemental As.
This inconsistency is due to the fact that the distribution of interatomic distances in the As-type structure for Bi is different from that in the As-type structure for As.

Moreover, some of the local minimum structures are assigned as prototype structures, such as C($P6_3mc$)-, simple-cubic (SC)-, $\beta$-Sn-, MnP-, Bi(II)-, black-P-, and Cs(HP)-type structures.
Among these structures, the SC-type \cite{jaggi1964struktur} and Bi(II)-type \cite{AkselrudHanflandmSchwarz+2003+447+448} structures were reported as high-pressure polymorphs in the literature.
The body-centered-cubic (BCC)-type structure, which is known as a high-pressure structure \cite{Schaufelberger1973}, is also found in the structure list obtained from the structure search using the MLPs.
However, it exhibits a relative energy value larger than $\theta =$ 50 meV/atom.
In the elemental Bi, Bi($I4/mcm$)-type \cite{degtyareva2001crystal}, Bi(III)-type \cite{degtyareva2001crystal}, Si($oS16$)-type \cite{chaimayo2012high}, and Sb($mP4$)-type \cite{kabalkina1970investigation} structures were also reported as high-pressure phases, but they are not found in the current structure list.

\subsubsection{Ga}

\begin{figure}[tbp]
\includegraphics[clip,width=\linewidth]{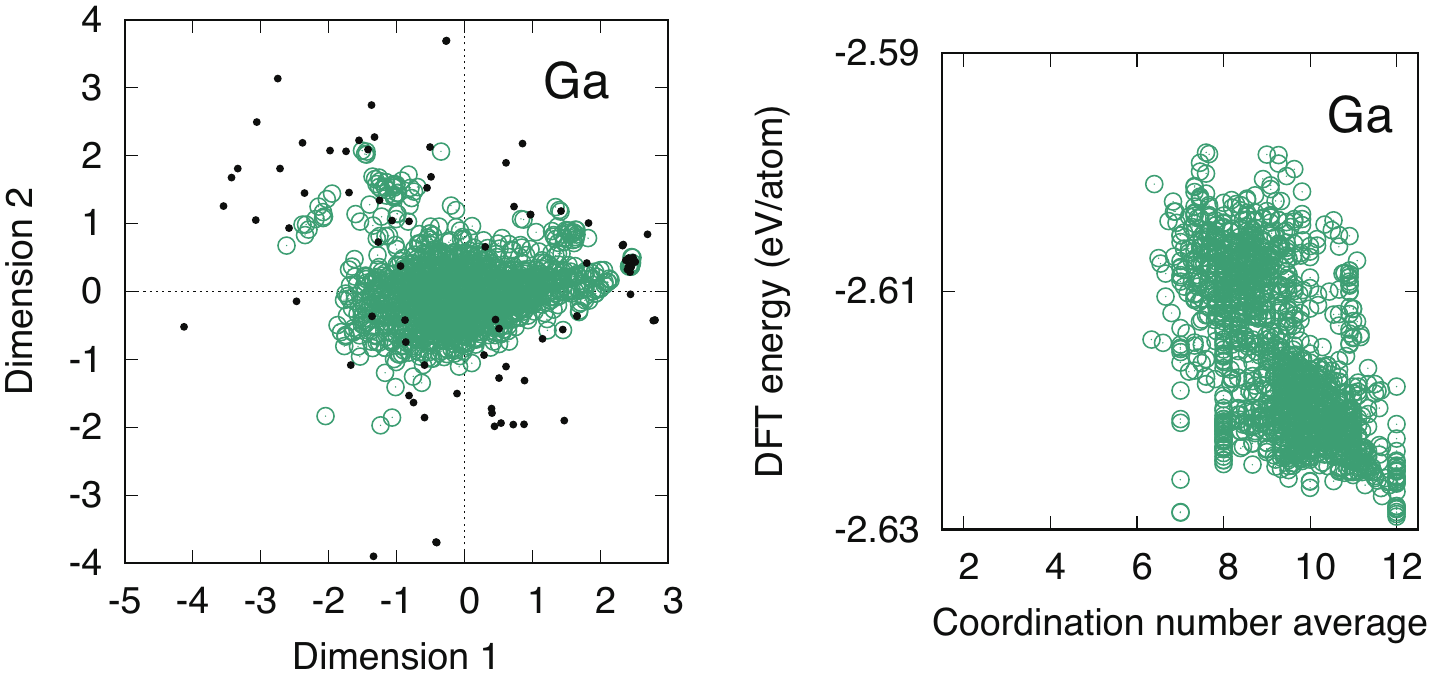}
\caption{
(a) Distribution of the local minimum structures and equilibrium prototype structures in Ga.
These structures are mapped into a two-dimensional plane using an MDS.
The open and closed circles show the local minimum structures and prototype structures, respectively.
(b) Average coordination numbers of the local minimum structures.
The cutoff radius is given as 1.2 times the nearest neighbor distance.
}
\label{mlp-go:Fig-structure-Ga}
\end{figure}

There are 1142 local minimum structures in the elemental Ga with relative energy values less than $\theta =$ 25 meV/atom. 
Among these structures, 616 structures are highly competitive in terms of energy, with relative energy values less than 15 meV/atom.
Figure \ref{mlp-go:Fig-structure-Ga} (a) shows the distribution of the local minimum structures and equilibrium prototype structures in Ga.
The local minimum structures are located inside the distribution of the prototype structures, while many of them are distant from the prototype structures.
Figure \ref{mlp-go:Fig-structure-Ga} (b) shows the distribution of average coordination numbers for the local minimum structures in Ga.
The global minimum structure has a coordination number of seven, while 12-coordinated local minimum structures similar to the FCC structure are also competing with the global minimum structure.
The average coordination numbers of the local minimum structures in Ga range from six to twelve.

\begin{table}[tbp]
\begin{ruledtabular}
\caption{
Global and local minimum structures in Ga.
The DFT relative energy $\Delta E$ is expressed in the unit of meV/atom.
$Z$ denotes the number of atoms included in the unitcell.
}
\label{mlp-go:Table-structure-Ga}
\begin{tabular}{ccccc}
Space group & ICSD-ID & Prototype & $Z$ & $\Delta E$ \\
\hline
$Cmce$          &  43388 & Ga(I)  & 8   & 0.0 \\ 
$P\bar1$        & (12174)& (In)  & 10  & 5.2 \\ 
$C2/m$          & (12174)& (In)  & 16  & 5.4 \\ 
$Cmcm$          &  $-$   & $-$   & 8   & 5.5 \\ 
$P\bar1$        & (12174)& (In)  & 12  & 5.5 \\ 
$Imma$          &  $-$   &  $-$  & 16  & 5.6 \\ 
$P\bar1$        & (12174)& (In)  & 12  & 5.8 \\ 
$P1$            & (12174)& (In)  & 12  & 5.8 \\ 
$C2/m$          & (54338)& (Pr)  & 24  & 5.9 \\ 
$C2/m$          & (54338)& (Pr)  & 18  & 6.1 \\ 
$P\bar1$        & (12174)& (In)  & 12  & 6.1 \\ 
$Cmcm$          &  $-$   &  $-$  & 24  & 6.9 \\ 
$R\bar3m$       & (12174)& (In)  & 3   & 7.3 \\ 
$Cmmm$          &  $-$   &  $-$  & 22  & 7.6 \\ 
$Cmcm$          &  $-$   &  $-$  & 4   & 7.6 \\ 
$Cm$            &  $-$   &  $-$  & 24  & 7.8 \\ 
$Immm$          &  $-$   &  $-$  & 22  & 7.8 \\ 
$Fm\bar3m$      & 20502  &  FCC  & 4   & 7.9 \\ 
$Cm$            &  $-$   &  $-$  & 16  & 8.0 \\ 
$Cm$            &  $-$   &  $-$  & 20  & 8.1 \\ 
$R\bar3m$       &  $-$   &  $-$  & 30  & 8.1 \\ 
$Fddd$          & 44866  &Pu($oF8$)& 8   & 8.1 \\ 
$C2/c$          &  $-$   &  $-$  & 20  & 8.2 \\ 
$Immm$          &  $-$   &  $-$  & 22  & 8.2 \\ 
$Cmmm$          &  $-$   &  $-$  & 18  & 8.2 \\ 
$P4_22_12$      &  $-$   &  $-$  & 8   & 8.3 \\ 
$R\bar3m$       &  $-$   &  $-$  & 27  & 8.3 \\ 
$Cm$            &(52496) & (Tb)  & 12  & 8.3 \\ 
$Cmce$          &  $-$   &  $-$  & 16  & 8.4 \\ 
$Cmcm$          &  $-$   &  $-$  & 16  & 8.4 \\ 
$Cm$            &  $-$   &  $-$  & 16  & 8.4 \\ 
\hline
$Cm$            &(43573) & (La)  & 16  & 8.7 \\ 
$P6_3/mmc$      & 43573  & La   & 4   & 9.5 \\ 
$Cmcm$          & 43539  &Ga($Cmcm$)& 4   & 9.6 \\ 
$P6_3/mmc$      & 52496  & Tb   & 6   & 9.8 \\ 
$C2/m$          & 2284 &O$_2$($mS4$)& 20  & 10.1 \\ 
$Pnma$          & 165995 & MnP      & 8   & 10.8 \\ 
$P6_3/mmc$      & 164724 & O$_2$    & 4   & 11.2 \\ 
$R3c$           & (12173)& (Ga(II)) & 18  & 11.7 \\ 
$P2_1/c$        & 189806 &Bi($P2_1/c$)& 8   & 12.3 \\ 
$R\bar3m$       & 52497  & Sm  & 9   & 12.9 \\ 
$I\bar43d$      & 109012 &Li($cI16$)& 16  & 15.0 \\ 
$P6_3/mmc$      & 1425   & HCP & 2   & 16.1 \\ 
$P3_221$        & 281124 & S  & 9   & 22.0 \\ 
$P2_12_12$      & 16817&H$_2$O(III,IX)& 12& 23.8 \\ 
\end{tabular}
\end{ruledtabular}
\end{table}

Table \ref{mlp-go:Table-structure-Ga} presents a list of the local minimum structures in the elemental Ga.
The global minimum structure is the Ga(I)-type structure with the space group of $Cmce$, which is the experimental structure observed at low temperatures \cite{+1962+293+300,young1991phase}.
One of the local minimum structures, exhibiting the fourth lowest energy value and the space group of $Cmcm$, corresponds to a metastable structure determined through first-principle prediction \cite{PhysRevB.80.045209}.

In the elemental Ga, there are numerous competing close-packed and open structures.
The close-packed structures include In-, face-centered-cubic (FCC)-, La-, and Sm-type structures, while the open structures with coordination numbers of seven and eight correspond to Ga(I)- and Ga($Cmcm$)-type structures. Intermediate structures between the open and close-packed structures, namely Pu($oF8$)-, Ga(II)-, and Li($cI16$)-types, are also found as local minimum structures.

Experimental studies reported Ga($Cmcm$)- \cite{persee.fr:bulmi_0037-9328_1961_num_84_3_5486}, In- (Ga(III)-) \cite{10.1063/1.435841}, Ga(II)- \cite{10.1063/1.435841}, and Ga($R\bar3m$)($\delta$Ga)-type \cite{https://doi.org/10.1107/S0567740873002530} structures.
The first three structures are present in the list of local minimum structures. Although the Ga(II)-type structure is not equivalent to but close to a local minimum structure with a different space group, $R3c$. The Ga($R\bar3m$)($\delta$Ga)-type structure requires at least 22 atoms to represent, which cannot be discovered in the current procedure.

\subsubsection{In}

\begin{figure}[tbp]
\includegraphics[clip,width=\linewidth]{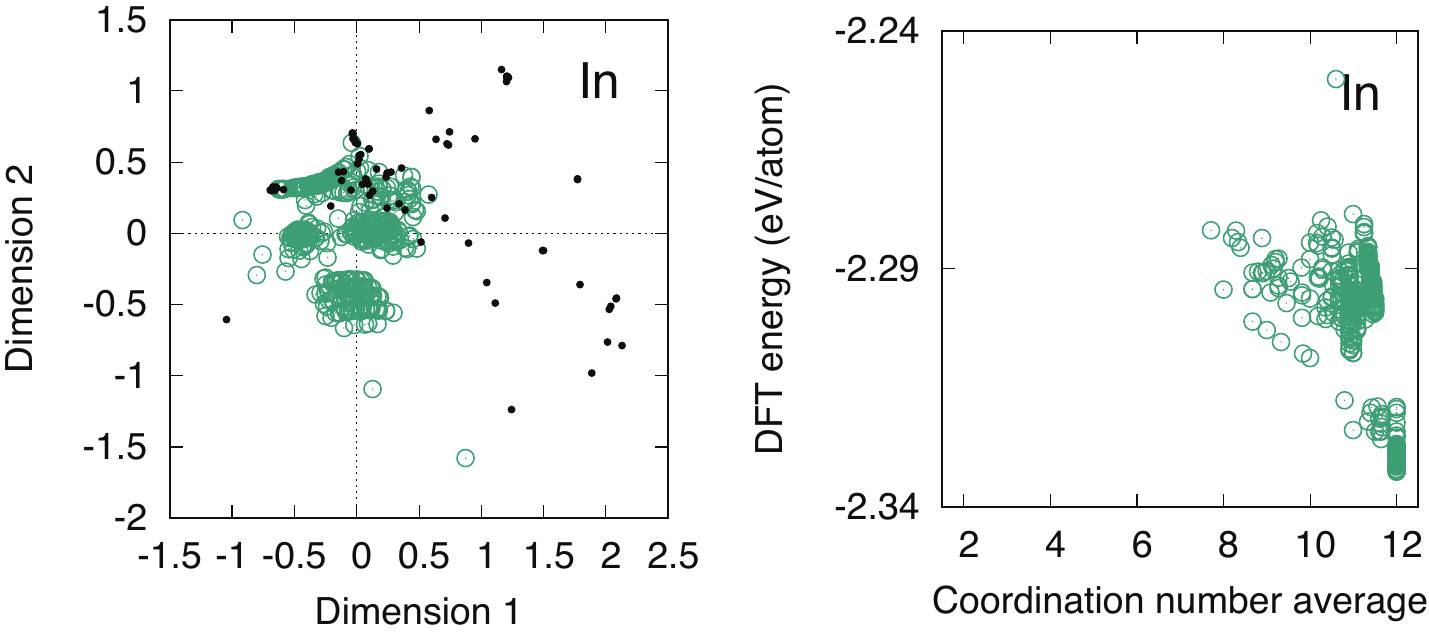}
\caption{
(a) Distribution of the local minimum structures and equilibrium prototype structures in In.
These structures are mapped into a two-dimensional plane using an MDS.
The open and closed circles show the local minimum structures and prototype structures, respectively.
(b) Average coordination numbers of the local minimum structures.
The cutoff radius is given as 1.2 times the nearest neighbor distance.
}
\label{mlp-go:Fig-structure-In}
\end{figure}

A total of 474 local minimum structures with relative energy values less than $\theta =$ 50 meV/atom have been discovered in the elemental In.
Among them, 99 structures are found for $\theta = 5$ meV/atom.
Figure \ref{mlp-go:Fig-structure-In} (a) shows the distribution of the local minimum structures and equilibrium prototype structures in the elemental In. Some local minimum structures are close to the equilibrium prototype structures, while many others are far from them.
Figure \ref{mlp-go:Fig-structure-In} (b) shows the distribution of average coordination numbers for the local minimum structures in the elemental In.
The global minimum structure is 12-coordinated, and the local minimum structures also exhibit large coordination numbers ranging from 10 to 12.

\begin{table}[tbp]
\begin{ruledtabular}
\caption{
Global and local minimum structures in In.
The DFT relative energy $\Delta E$ is expressed in the unit of meV/atom.
$Z$ denotes the number of atoms included in the unitcell.
}
\label{mlp-go:Table-structure-In}
\begin{tabular}{ccccc}
Space group & ICSD-ID & Prototype & $Z$ & $\Delta E$ \\
\hline
$C2/m$          & (12174)& (In)  & 20  & 0.0 \\
                & (57392)&(In($Fmmm$))&     &     \\
                & (41824)& (Ce)  &     &     \\ 
$I4/mmm$        &  12174 &  In   & 2   & 0.1 \\ 
$C2/m$          &  54338 &  Pr   & 4   & 0.4 \\ 
$C2/m$          & (12174)& (In)  & 16  & 0.4 \\ 
$C2/m$          &  41824 &  Ce   & 2   & 0.6 \\ 
$P\bar1$        & (41824)& (Ce)  & 1   & 1.0 \\ 
$Pmmn$          & (12174)& (In)  & 10  & 1.2 \\ 
$C2/m$          &  $-$   &  $-$  & 24  & 1.6 \\ 
$C2/m$          &  $-$   &  $-$  & 18  & 1.7 \\ 
$R\bar3m$       &  $-$   &  $-$  & 24  & 1.8 \\ 
$C2/m$          &  $-$   &  $-$  & 24  & 1.8 \\ 
$P6_3/mmc$      &  $-$   &  $-$  & 8   & 1.8 \\ 
$Cmcm$          &  $-$   &  $-$  & 20  & 1.9 \\ 
$R\bar3m$       &  $-$   &  $-$  & 27  & 1.9 \\ 
$P\bar6m2$      &  $-$   &  $-$  & 12  & 1.9 \\ 
$C2/m$          &  $-$   &  $-$  & 20  & 1.9 \\ 
$R\bar3m$       &  $-$   &  $-$  & 24  & 2.0 \\ 
$C2/m$          &  $-$   &  $-$  & 20  & 2.0 \\ 
$P\bar3m1$      &  $-$   &  $-$  & 9   & 2.1 \\ 
$C2/m$          &  $-$   &  $-$  & 22  & 2.1 \\ 
$C2/m$          &  $-$   &  $-$  & 20  & 2.2 \\ 
$R\bar3m$       &  $-$   &  $-$  & 18  & 2.2 \\ 
$P6_3/mmc$      & 43573  &  La   & 4   & 2.3 \\ 
$C2/m$          &  $-$   &  $-$  & 18  & 2.3 \\ 
$C2/m$          &  $-$   &  $-$  & 16  & 2.3 \\ 
$R\bar3m$       &  $-$   &  $-$  & 30  & 2.3 \\ 
$P\bar3m1$      &  $-$   &  $-$  & 7   & 2.4 \\ 
$R3m$           &  $-$   &  $-$  & 36  & 2.4 \\ 
$R3m$           &  $-$   &  $-$  & 33  & 2.4 \\ 
\hline
$P6_3/mmc$      & 52496  &  Tb   & 6   & 2.6 \\ 
$R\bar3m$       & 52497  &  Sm   & 9   & 5.3 \\ 
$P6_3/mmc$      & 1425   &  HCP  & 2   & 5.7 \\ 
$I\bar43d$      & 109012 &Li($cI16$)& 16  & 8.7 \\ 
$P1$            & (5248) & (BCC) & 11  & 10.3 \\ 
$P3_221$        & 281124 &  S    & 9   & 38.3 \\ 
\end{tabular}
\end{ruledtabular}
\end{table}

Table \ref{mlp-go:Table-structure-In} provides a list of the local minimum structures present in the elemental In.
The global minimum structure has the $C2/m$ space group, is 12-coordinated, and is almost identical to the In-type \cite{PhysRev.17.549} and In($Fmmm$)-type \cite{PhysRevB.47.8465} structures.
However, the high-pressure In($Fmmm$)-type structure, reported in the literature, is quite different from the global minimum structure in terms of the lattice constants.
The local minimum structure with the second lowest energy can be identified with the In-type structure, which is the experimental structure at room temperature \cite{PhysRev.17.549,young1991phase}.
Table \ref{mlp-go:Table-structure-In} shows that there are many local minimum structures with low relative energy values, including La-, Tb-, Sm-, and hexagonal-close-packed (HCP)-type structures.
In the literature, the In- and In($Fmmm$)-type structures are included in the ICSD. 
They are found in the list of local minimum structures.

\subsubsection{La}

\begin{figure}[tbp]
\includegraphics[clip,width=\linewidth]{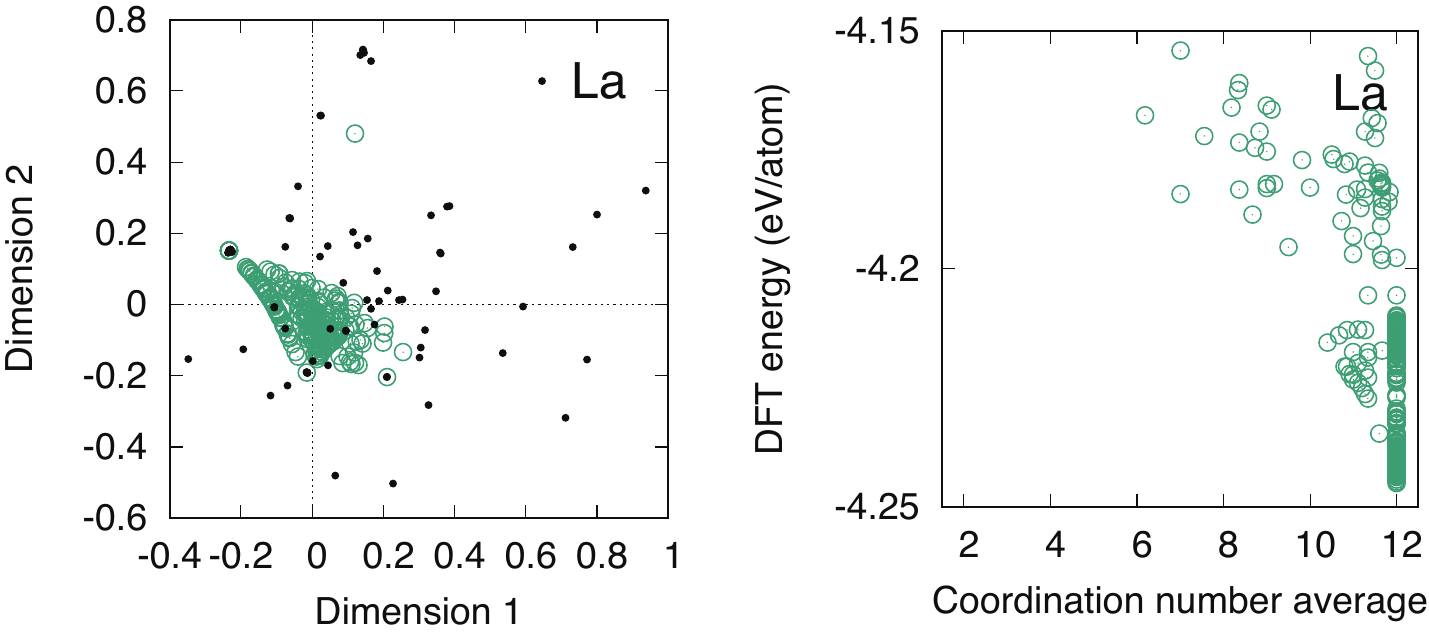}
\caption{
(a) Distribution of the local minimum structures and equilibrium prototype structures in La.
These structures are mapped into a two-dimensional plane using an MDS.
The open and closed circles show the local minimum structures and prototype structures, respectively.
(b) Average coordination numbers of the local minimum structures.
The cutoff radius is given as 1.2 times the nearest neighbor distance.
}
\label{mlp-go:Fig-structure-La}
\end{figure}

In the case of the elemental La, 253 local minimum structures have been identified, all of which have relative energy values less than $\theta =$ 75 meV/atom.
Among these structures, 73 structures have been discovered for $\theta = 15$ meV/atom.
Figure \ref{mlp-go:Fig-structure-La} (a) shows the distribution of the local minimum structures and equilibrium prototype structures in La.
Many local minimum structures are distributed around the equilibrium prototype structures.
Figure \ref{mlp-go:Fig-structure-La} (b) depicts the distribution of average coordination numbers for the local minimum structures in La.
The global minimum structure and many local minimum structures are 12-coordinated, while some local minimum structures have smaller coordination numbers ranging from six to eight.

\begin{table}[tbp]
\begin{ruledtabular}
\caption{
Global and local minimum structures in La.
The DFT relative energy $\Delta E$ is expressed in the unit of meV/atom.
$Z$ denotes the number of atoms included in the unitcell.
}
\label{mlp-go:Table-structure-La}
\begin{tabular}{ccccc}
Space group & ICSD-ID & Prototype & $Z$ & $\Delta E$ \\
\hline
$P6_3/mmc$      & 43573  &  La   & 4   & 0.0 \\ 
$R\bar3m$       &  $-$   &  $-$  & 27  & 0.3 \\ 
$R\bar3m$       &  $-$   &  $-$  & 15  & 0.7 \\ 
$P\bar6m2$      &  $-$   &  $-$  & 10  & 1.2 \\ 
$R3m$           &  $-$   &  $-$  & 33  & 1.3 \\ 
$R\bar3m$       &  $-$   &  $-$  & 33  & 1.4 \\ 
$R\bar3m$       &  $-$   &  $-$  & 18  & 1.6 \\ 
$P\bar3m1$      &  $-$   &  $-$  & 7   & 1.7 \\ 
$R\bar3m$       &  $-$   &  $-$  & 24  & 1.9 \\ 
$R\bar3m$       &  $-$   &  $-$  & 33  & 1.9 \\ 
$R3m$           &  $-$   &  $-$  & 36  & 2.1 \\ 
$R\bar3m$       &  $-$   &  $-$  & 21  & 2.3 \\ 
$R\bar3m$       &  $-$   &  $-$  & 27  & 2.3 \\ 
$P\bar3m1$      &  $-$   &  $-$  & 9   & 2.4 \\ 
$R\bar3m$       &  $-$   &  $-$  & 21  & 2.8 \\ 
$P6_3/mmc$      &  $-$   &  $-$  & 8   & 2.8 \\ 
$P6_3/mmc$      & 52496  &  Tb   & 6   & 2.9 \\ 
$P6_3/mmc$      &  $-$   &  $-$  & 10  & 2.9 \\ 
$P\bar3m1$      &  $-$   &  $-$  & 8   & 3.0 \\ 
$P3m1$          &  $-$   &  $-$  & 9   & 3.4 \\ 
$R3m$           &  $-$   &  $-$  & 33  & 3.4 \\ 
$R3m$           &  $-$   &  $-$  & 30  & 3.7 \\ 
$P3m1$          &  $-$   &  $-$  & 12  & 3.8 \\ 
$R3m$           &  $-$   &  $-$  & 24  & 4.1 \\ 
$R3m$           &  $-$   &  $-$  & 36  & 4.1 \\ 
$P3m1$          &  $-$   &  $-$  & 10  & 4.4 \\ 
$P3m1$          &  $-$   &  $-$  & 11  & 4.4 \\ 
$R3m$           &  $-$   &  $-$  & 33  & 4.6 \\ 
$Fm\bar3m$      & 20502  &  FCC  & 4   & 4.7 \\ 
$R3m$           &  $-$   &  $-$  & 27  & 4.8 \\ 
$P\bar3m1$      &  $-$   &  $-$  & 9   & 5.0 \\ 
\hline
$R\bar3m$       & 52497  &  Sm   & 9   & 6.2 \\ 
$P6_3/mmc$      &  1425  &  HCP  & 2   & 29.5 \\ 
$P6/mmm$        & 52521 &CaHg$_2$& 3   & 29.7 \\ 
\end{tabular}
\end{ruledtabular}
\end{table}

The local minimum structures in the elemental La are listed in Table \ref{mlp-go:Table-structure-La}. 
The 12-coordinated La-type structure is the global minimum structure, which is consistent with the experimental structure at low temperatures \cite{spedding1961high,young1991phase}. 
Additionally, several close-packed local minimum structures, such as the FCC-, Sm-, and HCP-type structures, have low energy values near the global minimum energy. 
Among these structures, the FCC-type structure is recognized as a high-temperature polymorph \cite{spedding1961high,young1991phase}.
Although the BCC-type structure is also known as a high-temperature structure \cite{spedding1961high,young1991phase}, it is not found in the structure list since it is dynamically unstable at zero temperature.
Moreover, the Sn($tI2$)-type structure \cite{nomura1977lanthanum} is not included in the structure list.

\subsubsection{P}

\begin{figure}[tbp]
\includegraphics[clip,width=\linewidth]{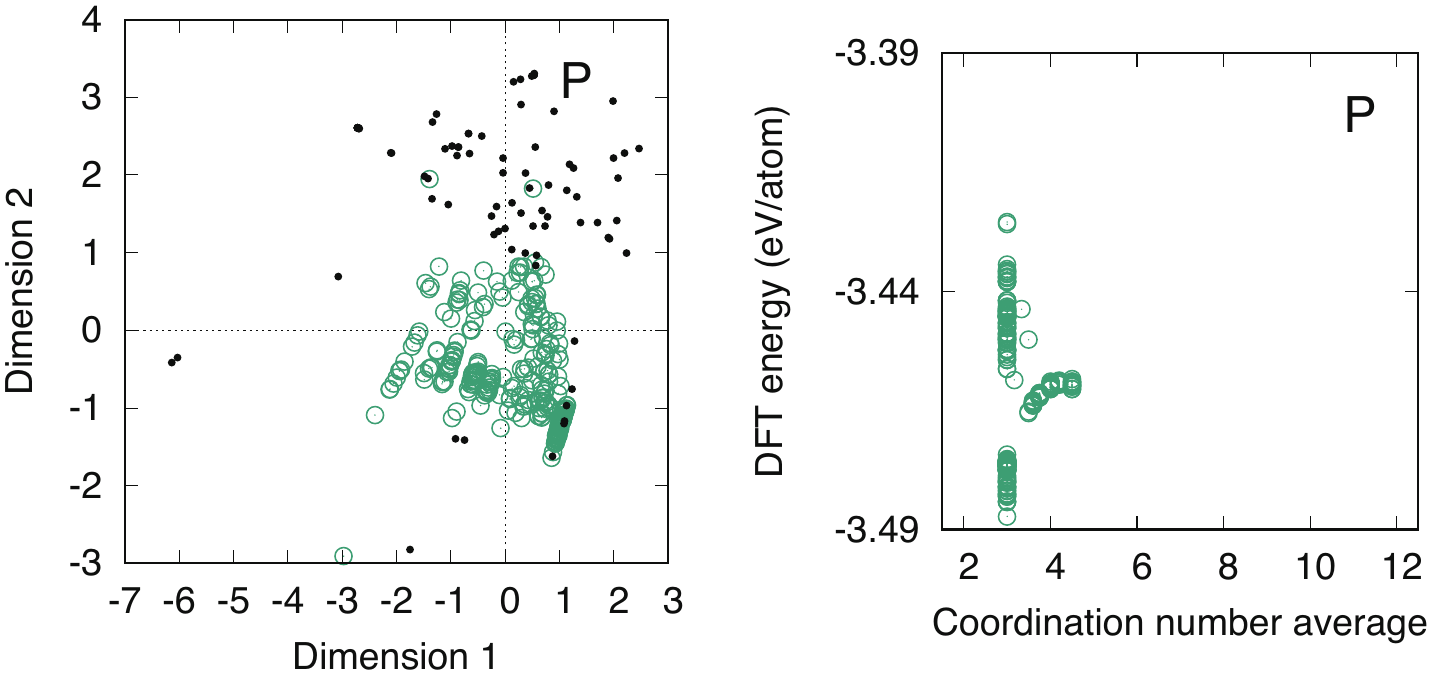}
\caption{
(a) Distribution of the local minimum structures and equilibrium prototype structures in P.
These structures are mapped into a two-dimensional plane using an MDS.
The open and closed circles show the local minimum structures and prototype structures, respectively.
(b) Average coordination numbers of the local minimum structures.
The cutoff radius is given as 1.2 times the nearest neighbor distance.
}
\label{mlp-go:Fig-structure-P}
\end{figure}

In this study, 276 local minimum structures with relative energy values less than $\theta =$ 75 meV/atom have been identified for the elemental P.
In particular, 112 structures are discovered for $\theta = 15$ meV/atom.
Figure \ref{mlp-go:Fig-structure-P} (a) shows the distribution of the local minimum structures and equilibrium prototype structures in P.
Most of the local minimum structures are located far from the equilibrium prototype structures. Figure \ref{mlp-go:Fig-structure-P} (b) presents the distribution of average coordination numbers for the local minimum structures in P. The global minimum structure and most of the local minimum structures have three-coordinated configurations.

\begin{table}[tbp]
\begin{ruledtabular}
\caption{
Global and local minimum structures in P.
The DFT relative energy $\Delta E$ is expressed in the unit of meV/atom.
$Z$ denotes the number of atoms included in the unitcell.
}
\label{mlp-go:Table-structure-P}
\begin{tabular}{ccccc}
Space group & ICSD-ID & Prototype & $Z$ & $\Delta E$ \\
\hline
$Cmce$          & 23836  &black-P& 8   & 0.0 \\ 
$Ibam$          &  $-$   &  $-$  & 16  & 3.0 \\ 
$C2/m$          &  $-$   &  $-$  & 20  & 3.1 \\ 
$P\bar1$        &  $-$   &  $-$  & 12  & 3.1 \\ 
$C2/m$          &  $-$   &  $-$  & 12  & 4.1 \\ 
$P\bar1$        &  $-$   &  $-$  & 12  & 4.1 \\ 
$P\bar1$        &  $-$   &  $-$  & 8   & 4.5 \\ 
$Cm$            &  $-$   &  $-$  & 24  & 4.6 \\ 
$C2/m$          &  $-$   &  $-$  & 24  & 4.7 \\ 
$C2/m$          &  $-$   &  $-$  & 24  & 4.8 \\ 
$P\bar1$        &  $-$   &  $-$  & 6   & 5.3 \\ 
$C2/m$          &  $-$   &  $-$  & 20  & 5.5 \\ 
$Cccm$          &  $-$   &  $-$  & 8   & 5.9 \\ 
$P2/m$          &  $-$   &  $-$  & 12  & 6.1 \\ 
$C2/m$          &  $-$   &  $-$  & 16  & 6.3 \\ 
$C2/m$          &  $-$   &  $-$  & 16  & 6.3 \\ 
$P2/m$          &  $-$   &  $-$  & 12  & 6.3 \\ 
$P\bar1$        &  $-$   &  $-$  & 12  & 7.2 \\ 
$Cm$            &  $-$   &  $-$  & 24  & 7.2 \\ 
$Pm$            &  $-$   &  $-$  & 12  & 7.3 \\ 
$C2/m$          &  $-$   &  $-$  & 20  & 7.4 \\ 
$C2/m$          &  $-$   &  $-$  & 20  & 7.7 \\ 
$P\bar1$        &  $-$   &  $-$  & 8   & 7.7 \\ 
$C2/m$          &  $-$   &  $-$  & 24  & 8.0 \\ 
\hline
$P\bar3m1$      & (1425) & (HCP) & 2   & 10.4 \\ 
$P6_3mc$        & 31170  &C($P6_3mc$)& 4   & 10.5 \\ 
$R\bar3m$       & 29123  &Graphite(3R)& 6   & 10.7 \\ 
$I2_13$         & 187643 &  N    & 8   & 35.2 \\ 
$R\bar3m$       & 9859   &  As   & 6   & 50.3 \\ 
\end{tabular}
\end{ruledtabular}
\end{table}

Table \ref{mlp-go:Table-structure-P} lists the local minimum structures in the elemental P.
The global minimum structure is the three-coordinated black-P-type structure, also known as the most stable experimental structure at room temperature \cite{brown1965refinement,young1991phase}.
The structure list includes many other three-coordinated local minimum structures such as the C($P6_3mc$)-, graphite(3R)-, and As-type structures. However, only the As-type structure is known as a high-pressure polymorph \cite{jamieson1963crystal,young1991phase}.
Several other experimental structures such as the SC- \cite{kikegawa1983x,young1991phase}, Ca(III)- \cite{PhysRevB.78.054120}, P($Amm2$)-types \cite{PhysRevB.78.054120} have also been reported as high-pressure polymorphs in the literature.
Among these structures, only the SC-type structure \cite{kikegawa1983x,young1991phase} is included in the list of local minimum structures obtained using the structure search using the MLP, but it shows an energy value larger than the threshold.

\subsubsection{Sb}

\begin{figure}[tbp]
\includegraphics[clip,width=\linewidth]{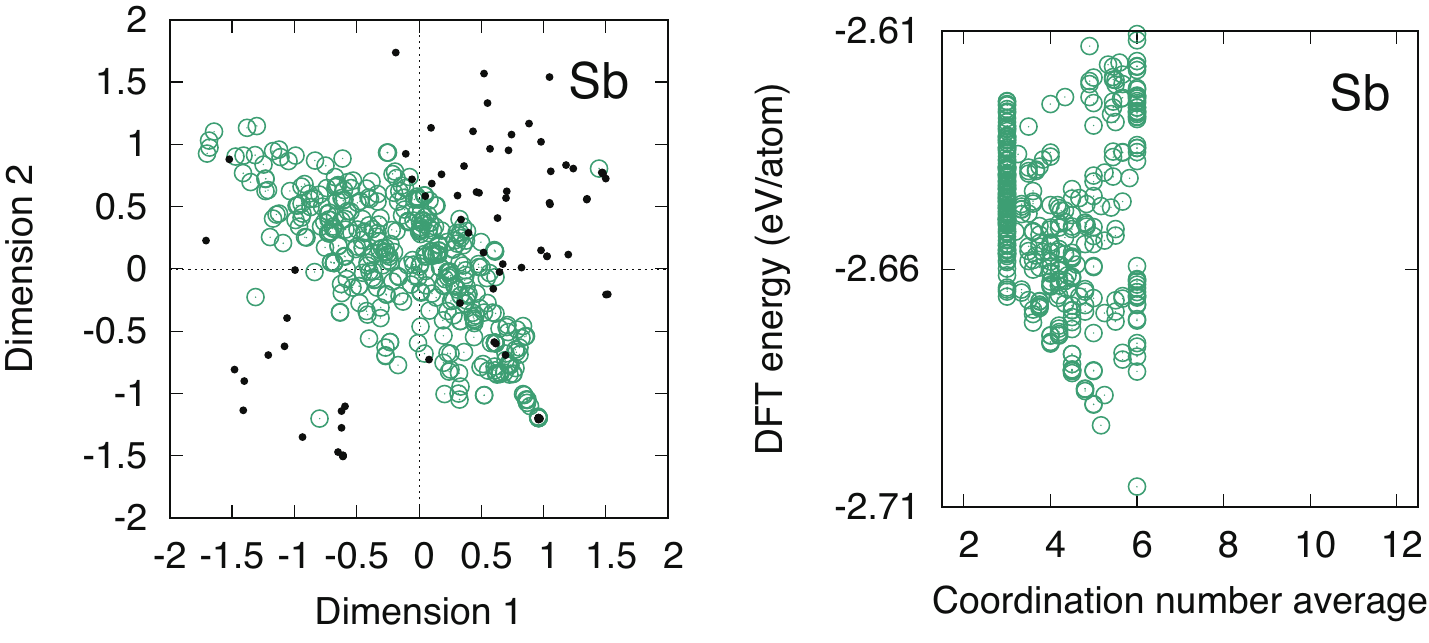}
\caption{
(a) Distribution of the local minimum structures and equilibrium prototype structures in Sb.
These structures are mapped into a two-dimensional plane using an MDS.
The open and closed circles show the local minimum structures and prototype structures, respectively.
(b) Average coordination numbers of the local minimum structures.
The cutoff radius is given as 1.2 times the nearest neighbor distance.
}
\label{mlp-go:Fig-structure-Sb}
\end{figure}

In the elemental Sb, 302 structures in the elemental Sb that have relative energy values less than $\theta =$ 75 meV/atom have been discovered. 
However, for $\theta = 15$ and 30 meV/atom, only two and 24 structures have been found, respectively.
Figure \ref{mlp-go:Fig-structure-Sb} (a) shows the distribution of the local minimum structures and equilibrium prototype structures in Sb.
Most of the local minimum structures are far from the equilibrium prototype structures.
Figure \ref{mlp-go:Fig-structure-Sb} (b) shows the distribution of average coordination numbers for the local minimum structures in Sb.
The elemental Sb prefers open structures, and the global minimum structure is six-coordinated.
The coordination numbers of the local minimum structures range from three to six.

\begin{table}[tbp]
\begin{ruledtabular}
\caption{
Global and local minimum structures in Sb.
The DFT relative energy $\Delta E$ is expressed in the unit of meV/atom.
$Z$ denotes the number of atoms included in the unitcell.
}
\label{mlp-go:Table-structure-Sb}
\begin{tabular}{ccccc}
Space group & ICSD-ID & Prototype & $Z$ & $\Delta E$ \\
\hline
$R\bar3m$       &  9859  &  As   & 6   & 0.0 \\ 
$Pm$            &  $-$   &  $-$  & 12  & 12.9 \\ 
$R\bar3m$       &  $-$   &  $-$  & 36  & 17.3 \\ 
$P6_3mc$        &  $-$   &  $-$  & 12  & 17.3 \\ 
$P\bar3m1$      &  $-$   &  $-$  & 8   & 19.3 \\ 
$P\bar3m1$      &  $-$   &  $-$  & 10  & 20.1 \\ 
$R\bar3m$       &  $-$   &  $-$  & 30  & 20.6 \\ 
$P3m1$          &  $-$   &  $-$  & 10  & 20.6 \\ 
$P6_3mc$        &  $-$   &  $-$  & 8   & 23.9 \\ 
$Imma$          &  $-$   &  $-$  & 24  & 24.3 \\ 
$R\bar3m$       &  $-$   &  $-$  & 24  & 24.3 \\ 
$R\bar3m$       &  $-$   &  $-$  & 18  & 24.4 \\ 
$R\bar3m$       &  $-$   &  $-$  & 24  & 24.4 \\ 
$P1$            &  $-$   &  $-$  & 8   & 24.4 \\ 
$C2/m$          &  $-$   &  $-$  & 12  & 24.5 \\ 
$P1$            &  $-$   &  $-$  & 8   & 24.6 \\ 
$R3m$           &  $-$   &  $-$  & 24  & 24.7 \\ 
$P\bar3m1$      &  $-$   &  $-$  & 8   & 25.8 \\ 
$Cmce$          &  $-$   &  $-$  & 24  & 26.7 \\ 
$Pmma$          &  $-$   &  $-$  & 10  & 27.4 \\ 
$P3m1$          &  $-$   &  $-$  & 12  & 27.8 \\ 
$R3m$           &  $-$   &  $-$  & 36  & 27.9 \\ 
$Cccm$          &  $-$   &  $-$  & 24  & 28.1 \\ 
$C2/m$          &  $-$   &  $-$  & 16  & 29.4 \\ 
\hline
$P6_3mc$        & 31170  &C($P6_3mc$)& 4  & 40.0 \\ 
$Pm\bar3m$      & 43211  & SC    & 1 & 46.4 \\ 
$Cmce$          & 23836  &black-P& 8 & 47.2 \\ 
$Pnma$          & 165995 &  MnP  & 8 & 51.0 \\ 
\end{tabular}
\end{ruledtabular}
\end{table}

Table \ref{mlp-go:Table-structure-Sb} provides a list of the local minimum structures observed in the elemental Sb.
The global minimum structure is the six-coordinated As-type structure, which is also the experimental structure at low temperatures \cite{barrett1963crystal,young1991phase}.
In addition, the coordination numbers for all atoms in the As-type structure are obtained as six, similar to the elemental Bi.
Other structures such as the C($P6_3mc$)-, SC-, black-P- and MnP-type structures are also identified. The SC-type structure is the only known high-pressure polymorph \cite{vereschchagin1965phase} among these structures.
Additionally, the BCC- \cite{aoki1983new}, HCP- \cite{vereschchagin1965phase}, Sb($mP4$)- \cite{kabalkina1970investigation}, Bi($I4/mcm$)- \cite{PhysRevB.67.214101}, and Bi(III)-type structures \cite{PhysRevB.67.214101} are known as high-pressure polymorphs and have been experimentally observed.
Among these structures, the BCC-, HCP-, Sb($mP4$)-, and Bi($I4/mcm$)-type structures are included in the list of local minimum structures obtained using the random structure search using the MLP.

\subsubsection{Sn}

\begin{figure}[tbp]
\includegraphics[clip,width=\linewidth]{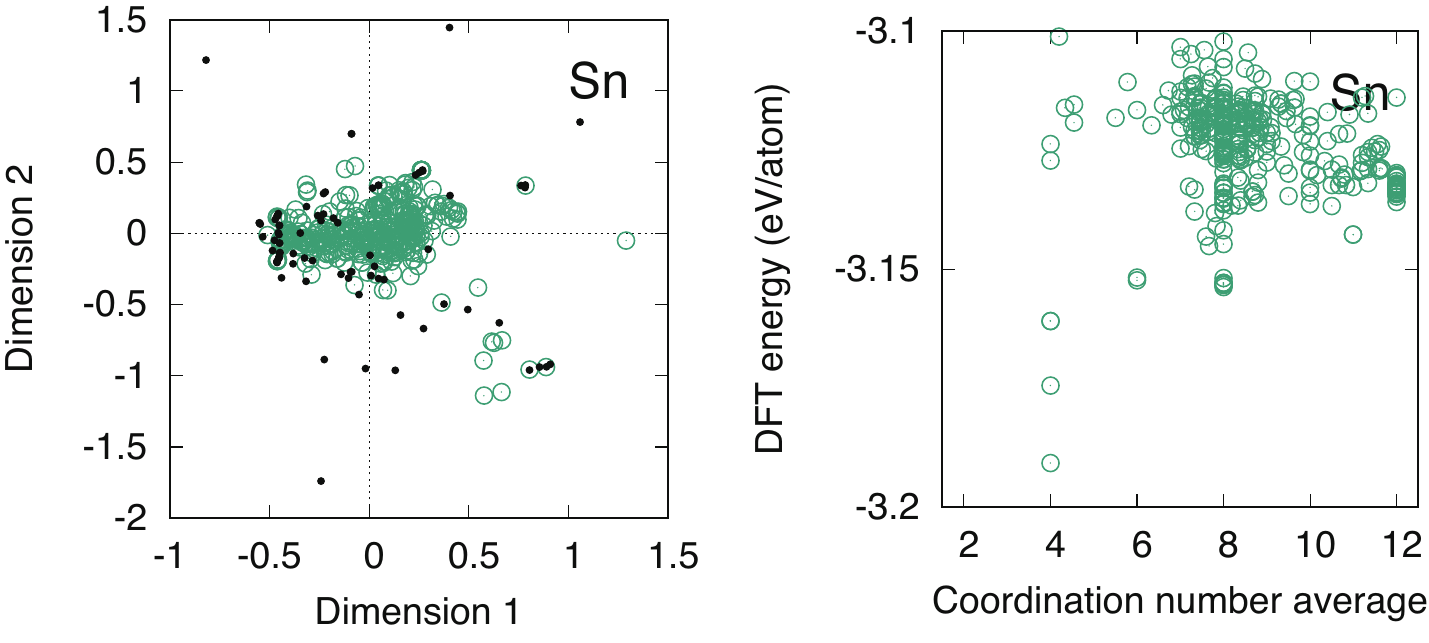}
\caption{
(a) Distribution of the local minimum structures and equilibrium prototype structures in Sn.
These structures are mapped into a two-dimensional plane using an MDS.
The open and closed circles show the local minimum structures and prototype structures, respectively.
(b) Average coordination numbers of the local minimum structures.
The cutoff radius is given as 1.2 times the nearest neighbor distance.
}
\label{mlp-go:Fig-structure-Sn}
\end{figure}

There are 310 local minimum structures in the elemental Sn, each with relative energy values lower than $\theta =$ 75 meV/atom.
However, only one, four, and thirteen structures are discovered for $\theta =$ 15, 30, and 45 meV/atom, respectively. 
Figure \ref{mlp-go:Fig-structure-Sn} (a) illustrates the distribution of the local minimum structures and equilibrium prototype structures in Sn. 
Most of the local minimum structures are located around the equilibrium prototype structures. 
Figure \ref{mlp-go:Fig-structure-Sn} (b) shows the distribution of average coordination numbers for the local minimum structures in Sn. 
The global minimum structure and the local minimum structures with low energy values are four-coordinated, while the coordination number of the local minimum structures ranges from four to twelve.

\begin{table}[tbp]
\begin{ruledtabular}
\caption{
Global and local minimum structures in Sn.
The DFT relative energy $\Delta E$ is expressed in the unit of meV/atom.
$Z$ denotes the number of atoms included in the unitcell.
}
\label{mlp-go:Table-structure-Sn}
\begin{tabular}{ccccc}
Space group & ICSD-ID & Prototype & $Z$ & $\Delta E$ \\
\hline
$Fd\bar3m$      &28857&Diamond-C($cF8$)  & 8 & 0.0 \\ 
$P6_3/mmc$      &30101&Wurtzite-ZnS(2H)& 4 & 16.3 \\ 
$P6_122$        &  $-$   &  $-$  & 6   & 29.8 \\ 
$P6_522$        &  $-$   &  $-$  & 6   & 29.8 \\ 
$P6/mmm$        &52456&Ca$_{.15}$Sn$_{.85}$& 1   & 37.0 \\ 
$Cmmm$          &(52456)&(Ca$_{.15}$Sn$_{.85}$)& 10&37.4 \\ 
$Cmm2$          &(52456)&(Ca$_{.15}$Sn$_{.85}$)& 22&37.5 \\ 
$P\bar3m1$      &(52456)&(Ca$_{.15}$Sn$_{.85}$)& 3 &37.8 \\ 
$Pmma$          &(52456)&(Ca$_{.15}$Sn$_{.85}$)& 8 &37.8 \\ 
$Pmna$          &(52456)&(Ca$_{.15}$Sn$_{.85}$)& 10&37.8 \\ 
$P1$            &(40037) &($\beta$-Sn)& 8 & 38.4 \\ 
$Pmc2_1$        &(40037) &($\beta$-Sn)& 8 & 38.9 \\ 
$I4_1/amd$      & 40037  & $\beta$-Sn & 4 & 39.0 \\ 
$Fdd2$          &  $-$   &  $-$  & 48  & 45.6 \\ 
$Cm$            &  $-$   &  $-$  & 24  & 46.0 \\ 
$P2_1/c$        &  $-$   &  $-$  & 10  & 47.6 \\ 
$I\bar43d$      &109012 &Li($cI16$)& 16  & 48.0 \\ 
$I\bar43d$      &109012 &Li($cI16$)& 16  & 48.0 \\ 
$C2/c$          &  $-$   &  $-$  & 16  & 48.9 \\ 
$C2/m$          &  $-$   &  $-$  & 24  & 49.8 \\ 
\hline
$Cmce$          & 89414 &Si($oS16$) & 16  & 54.0 \\ 
$Cm$            & (1425)& (HCP)   & 24  & 54.8 \\ 
$Cmcm$          & (12174)& (In)   & 4   & 56.0 \\ 
$I4/mmm$        & 12174  &  In    & 2   & 56.4 \\ 
                & (236711)& (I($Immm$)) & & \\ 
$I4/mmm$        & 24622  &Sn($tI2$)  & 2   & 56.8 \\ 
$I4/mmm$        & 181908 &Si($I4/mmm$)& 8   & 67.1 \\ 
$Pm\bar3m$         & 43211  &SC& 1   & 74.2 \\ 
\end{tabular}
\end{ruledtabular}
\end{table}

The local minimum structures in the elemental Sn are listed in Table \ref{mlp-go:Table-structure-Sn}.
The global minimum structure is the four-coordinated diamond-type structure, which is the experimental structure at low temperatures \cite{swanson1953standard,young1991phase}.
The local minimum structures also include the wurtzite-, $\beta$-Sn-, Li($cI16$)-, Si($oS16$)-, In-, and Sn($tI2$)-type structures.
Experimentally, the $\beta$-Sn- \cite{lee1954lattice,young1991phase}, Sn($tI2$)- \cite{barnett1966x}, I($Immm$)- \cite{PhysRevB.88.104104}, and BCC-type \cite{PhysRevB.88.104104} structures have been reported as high-pressure polymorphs, and the $\beta$-Sn-type structure is also known as the high-temperature structure \cite{young1991phase}.
Among these structures, the $\beta$-Sn-, Sn($tI2$)-, and I($Immm$)-type structures are included in the current structure list.

\subsubsection{Te}

\begin{figure}[tbp]
\includegraphics[clip,width=\linewidth]{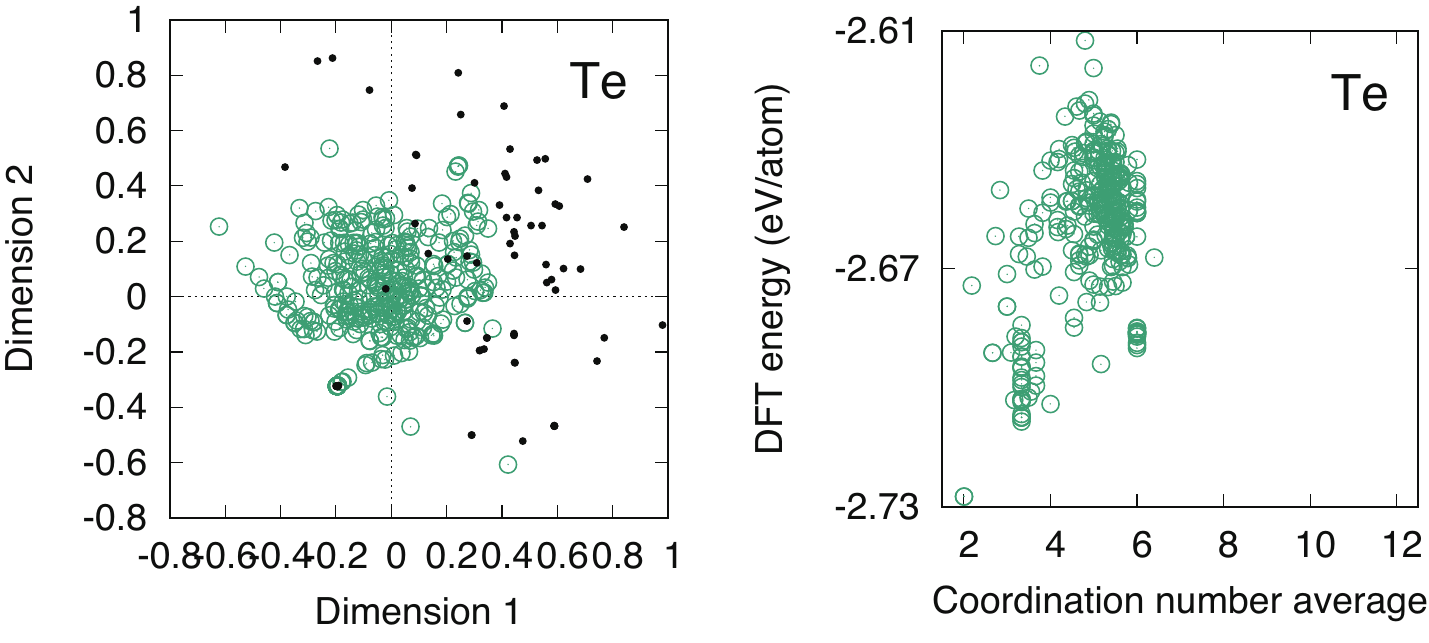}
\caption{
(a) Distribution of the local minimum structures and equilibrium prototype structures in Te.
These structures are mapped into a two-dimensional plane using an MDS.
The open and closed circles show the local minimum structures and prototype structures, respectively.
(b) Average coordination numbers of the local minimum structures.
The cutoff radius is given as 1.2 times the nearest neighbor distance.
}
\label{mlp-go:Fig-structure-Te}
\end{figure}

In the elemental Te, 339 local minimum structures with relative energy values less than $\theta =$ 100 meV/atom have been discovered.
On the other hand, only seven structures are discovered for $\theta =$ 20 meV/atom.
Figure \ref{mlp-go:Fig-structure-Te} (a) shows the distribution of the local minimum structures and equilibrium prototype structures in Te.
Most of the local minimum structures are far from the equilibrium prototype structures.
Figure \ref{mlp-go:Fig-structure-Te} (b) shows the distribution of average coordination numbers for the local minimum structures in Te.
The global minimum structure is two-coordinated, and the coordination numbers of the local minimum structures range from two to six.

\begin{table}[tbp]
\begin{ruledtabular}
\caption{
Global and local minimum structures in Te.
The DFT relative energy $\Delta E$ is expressed in the unit of meV/atom.
$Z$ denotes the number of atoms included in the unitcell.
}
\label{mlp-go:Table-structure-Te}
\begin{tabular}{ccccc}
Space group & ICSD-ID & Prototype & $Z$ & $\Delta E$ \\
\hline
$P3_121$        & 23059  &$\gamma$-Se& 3   & 0.0 \\ 
$P3_221$        & 653048 &  Te   & 3   & 0.0 \\ 
$P2$            &  $-$   &  $-$  & 9   & 18.9 \\ 
$C2/c$          &(23059)&($\gamma$-Se)& 12& 19.9 \\ 
$P2_1/c$        &(23059)&($\gamma$-Se)& 12& 19.9 \\ 
$P\bar1$        &(23059)&($\gamma$-Se)& 9 & 19.9 \\ 
$C2/c$          &(23059)&($\gamma$-Se)& 12& 20.0 \\ 
$P2$            &  $-$   &  $-$  & 12  & 20.7 \\ 
$R\bar3$        & 27495  &  S$_6$   & 18  & 23.4 \\ 
$Pm$            &  $-$   &  $-$  & 6   & 23.8 \\ 
$P1$            &  $-$   &  $-$  & 12  & 24.3 \\ 
$P2_1/c$        &  $-$   &  $-$  & 12  & 24.3 \\ 
$C2/m$          &  $-$   &  $-$  & 18  & 24.4 \\ 
$P2_12_12$      &  $-$   &  $-$  & 6   & 24.5 \\ 
$P2$            &  $-$   &  $-$  & 12  & 24.5 \\ 
$C2/m$          &  $-$   &  $-$  & 24  & 25.0 \\ 
$P2/m$          &  $-$   &  $-$  & 9   & 26.5 \\ 
$C2/m$          &  $-$   &  $-$  & 24  & 27.5 \\ 
$C2/m$          &  $-$   &  $-$  & 12  & 28.0 \\ 
$P1$            &  $-$   &  $-$  & 12  & 28.4 \\ 
$C2$            &  $-$   &  $-$  & 18  & 29.4 \\ 
$P2/m$          &  $-$   &  $-$  & 9   & 30.0 \\ 
\hline
$R\bar3m$       &(43211) & (SC)  & 3   & 40.8 \\ 
\end{tabular}
\end{ruledtabular}
\end{table}

Table \ref{mlp-go:Table-structure-Te} shows the local minimum structures in the elemental Te.
The global minimum structures are the two-coordinated $\gamma$-Se-type chiral structures, which are the experimental structures at low temperatures \cite{doi:10.1080/14786442408634511,young1991phase}.
However, most of the local minimum structures do not have assigned prototype structures.
In the literature, As- \cite{aoki1980crystal}, Sb$_2$Te$_3$- \cite{takumi2002x}, GaSb- \cite{vezzoli1971proposed}, Te($mP4$)- \cite{aoki1980crystal}, and Po($hR1$)-types \cite{aoki1980crystal} are known as high-pressure structures.
The first four structures are included in the list of local minimum structures obtained through the random structure search using MLP, showing energy values greater than the threshold.

\section{Conclusion}
\label{mlp-go:Section-conclusion}
This study has developed an efficient iterative procedure utilizing polynomial MLPs to enumerate both globally stable and metastable structures with high accuracy. 
The procedure involves updating the MLP and conducting a random structure search using the updated MLP.
This approach has enabled the creation of polynomial MLP models that are both reliable and efficient for structure enumeration. 

The current procedure has been applied to nine different elemental systems: As, Bi, Ga, In, La, P, Sb, Sn, and Te. In each of these systems, metastable structures that compete energetically with the globally stable structures have been identified. 
The study demonstrates the importance of exhaustively enumerating metastable structures with low energy values to achieve a comprehensive search for globally stable structures in such systems. 
By employing this procedure, globally stable structures consistent with experimental data at low temperatures have been identified. 
Additionally, many previously unknown metastable structures that compete with the globally stable structures have been discovered, as well as several metastable structures that correspond to experimental observations and prototype structures reported in other systems.
Therefore, this procedure will be valuable for conducting structure searches in complex systems characterized by potential energy surfaces with numerous local minima.

\begin{acknowledgments}
This work was supported by a Grant-in-Aid for Scientific Research (B) (Grant Number 22H01756) and a Grant-in-Aid for Scientific Research on Innovative Areas (Grant Number 19H05787) from the Japan Society for the Promotion of Science (JSPS).
\end{acknowledgments}

\bibliography{mlp-go}

\begin{thebibliography}{89}%
\makeatletter
\providecommand \@ifxundefined [1]{%
 \@ifx{#1\undefined}
}%
\providecommand \@ifnum [1]{%
 \ifnum #1\expandafter \@firstoftwo
 \else \expandafter \@secondoftwo
 \fi
}%
\providecommand \@ifx [1]{%
 \ifx #1\expandafter \@firstoftwo
 \else \expandafter \@secondoftwo
 \fi
}%
\providecommand \natexlab [1]{#1}%
\providecommand \enquote  [1]{``#1''}%
\providecommand \bibnamefont  [1]{#1}%
\providecommand \bibfnamefont [1]{#1}%
\providecommand \citenamefont [1]{#1}%
\providecommand \href@noop [0]{\@secondoftwo}%
\providecommand \href [0]{\begingroup \@sanitize@url \@href}%
\providecommand \@href[1]{\@@startlink{#1}\@@href}%
\providecommand \@@href[1]{\endgroup#1\@@endlink}%
\providecommand \@sanitize@url [0]{\catcode `\\12\catcode `\$12\catcode
  `\&12\catcode `\#12\catcode `\^12\catcode `\_12\catcode `\%12\relax}%
\providecommand \@@startlink[1]{}%
\providecommand \@@endlink[0]{}%
\providecommand \url  [0]{\begingroup\@sanitize@url \@url }%
\providecommand \@url [1]{\endgroup\@href {#1}{\urlprefix }}%
\providecommand \urlprefix  [0]{URL }%
\providecommand \Eprint [0]{\href }%
\providecommand \doibase [0]{http://dx.doi.org/}%
\providecommand \selectlanguage [0]{\@gobble}%
\providecommand \bibinfo  [0]{\@secondoftwo}%
\providecommand \bibfield  [0]{\@secondoftwo}%
\providecommand \translation [1]{[#1]}%
\providecommand \BibitemOpen [0]{}%
\providecommand \bibitemStop [0]{}%
\providecommand \bibitemNoStop [0]{.\EOS\space}%
\providecommand \EOS [0]{\spacefactor3000\relax}%
\providecommand \BibitemShut  [1]{\csname bibitem#1\endcsname}%
\let\auto@bib@innerbib\@empty
\bibitem [{\citenamefont {Lorenz}\ \emph {et~al.}(2004)\citenamefont {Lorenz},
  \citenamefont {Gro{\ss}},\ and\ \citenamefont {Scheffler}}]{Lorenz2004210}%
  \BibitemOpen
  \bibfield  {author} {\bibinfo {author} {\bibfnamefont {S.}~\bibnamefont
  {Lorenz}}, \bibinfo {author} {\bibfnamefont {A.}~\bibnamefont {Gro{\ss}}}, \
  and\ \bibinfo {author} {\bibfnamefont {M.}~\bibnamefont {Scheffler}},\ }\href
  {\doibase http://dx.doi.org/10.1016/j.cplett.2004.07.076} {\bibfield
  {journal} {\bibinfo  {journal} {Chem. Phys. Lett.}\ }\textbf {\bibinfo
  {volume} {395}},\ \bibinfo {pages} {210 } (\bibinfo {year}
  {2004})}\BibitemShut {NoStop}%
\bibitem [{\citenamefont {Behler}\ and\ \citenamefont
  {Parrinello}(2007)}]{behler2007generalized}%
  \BibitemOpen
  \bibfield  {author} {\bibinfo {author} {\bibfnamefont {J.}~\bibnamefont
  {Behler}}\ and\ \bibinfo {author} {\bibfnamefont {M.}~\bibnamefont
  {Parrinello}},\ }\href@noop {} {\bibfield  {journal} {\bibinfo  {journal}
  {Phys. Rev. Lett.}\ }\textbf {\bibinfo {volume} {98}},\ \bibinfo {pages}
  {146401} (\bibinfo {year} {2007})}\BibitemShut {NoStop}%
\bibitem [{\citenamefont {Behler}(2011)}]{behler2011atom}%
  \BibitemOpen
  \bibfield  {author} {\bibinfo {author} {\bibfnamefont {J.}~\bibnamefont
  {Behler}},\ }\href@noop {} {\bibfield  {journal} {\bibinfo  {journal} {J.
  Chem. Phys.}\ }\textbf {\bibinfo {volume} {134}},\ \bibinfo {pages} {074106}
  (\bibinfo {year} {2011})}\BibitemShut {NoStop}%
\bibitem [{\citenamefont {Han}\ \emph {et~al.}(2018)\citenamefont {Han},
  \citenamefont {Zhang}, \citenamefont {Car},\ and\ \citenamefont
  {Weinan}}]{han2017deep}%
  \BibitemOpen
  \bibfield  {author} {\bibinfo {author} {\bibfnamefont {J.}~\bibnamefont
  {Han}}, \bibinfo {author} {\bibfnamefont {L.}~\bibnamefont {Zhang}}, \bibinfo
  {author} {\bibfnamefont {R.}~\bibnamefont {Car}}, \ and\ \bibinfo {author}
  {\bibfnamefont {E.}~\bibnamefont {Weinan}},\ }\href@noop {} {\bibfield
  {journal} {\bibinfo  {journal} {Commun. Comput. Phys.}\ }\textbf {\bibinfo
  {volume} {23}},\ \bibinfo {pages} {629 } (\bibinfo {year}
  {2018})}\BibitemShut {NoStop}%
\bibitem [{\citenamefont {Artrith}\ and\ \citenamefont
  {Urban}(2016)}]{258c531ae5de4f5699e2eec2de51c84f}%
  \BibitemOpen
  \bibfield  {author} {\bibinfo {author} {\bibfnamefont {N.}~\bibnamefont
  {Artrith}}\ and\ \bibinfo {author} {\bibfnamefont {A.}~\bibnamefont
  {Urban}},\ }\href {\doibase 10.1016/j.commatsci.2015.11.047} {\bibfield
  {journal} {\bibinfo  {journal} {Comput. Mater. Sci.}\ }\textbf {\bibinfo
  {volume} {114}},\ \bibinfo {pages} {135} (\bibinfo {year}
  {2016})}\BibitemShut {NoStop}%
\bibitem [{\citenamefont {Artrith}\ \emph {et~al.}(2017)\citenamefont
  {Artrith}, \citenamefont {Urban},\ and\ \citenamefont
  {Ceder}}]{PhysRevB.96.014112}%
  \BibitemOpen
  \bibfield  {author} {\bibinfo {author} {\bibfnamefont {N.}~\bibnamefont
  {Artrith}}, \bibinfo {author} {\bibfnamefont {A.}~\bibnamefont {Urban}}, \
  and\ \bibinfo {author} {\bibfnamefont {G.}~\bibnamefont {Ceder}},\ }\href
  {\doibase 10.1103/PhysRevB.96.014112} {\bibfield  {journal} {\bibinfo
  {journal} {Phys. Rev. B}\ }\textbf {\bibinfo {volume} {96}},\ \bibinfo
  {pages} {014112} (\bibinfo {year} {2017})}\BibitemShut {NoStop}%
\bibitem [{\citenamefont {Bart{\'o}k}\ \emph {et~al.}(2010)\citenamefont
  {Bart{\'o}k}, \citenamefont {Payne}, \citenamefont {Kondor},\ and\
  \citenamefont {Cs{\'a}nyi}}]{bartok2010gaussian}%
  \BibitemOpen
  \bibfield  {author} {\bibinfo {author} {\bibfnamefont {A.~P.}\ \bibnamefont
  {Bart{\'o}k}}, \bibinfo {author} {\bibfnamefont {M.~C.}\ \bibnamefont
  {Payne}}, \bibinfo {author} {\bibfnamefont {R.}~\bibnamefont {Kondor}}, \
  and\ \bibinfo {author} {\bibfnamefont {G.}~\bibnamefont {Cs{\'a}nyi}},\
  }\href@noop {} {\bibfield  {journal} {\bibinfo  {journal} {Phys. Rev. Lett.}\
  }\textbf {\bibinfo {volume} {104}},\ \bibinfo {pages} {136403} (\bibinfo
  {year} {2010})}\BibitemShut {NoStop}%
\bibitem [{\citenamefont {Szlachta}\ \emph {et~al.}(2014)\citenamefont
  {Szlachta}, \citenamefont {Bart\'ok},\ and\ \citenamefont
  {Cs\'anyi}}]{PhysRevB.90.104108}%
  \BibitemOpen
  \bibfield  {author} {\bibinfo {author} {\bibfnamefont {W.~J.}\ \bibnamefont
  {Szlachta}}, \bibinfo {author} {\bibfnamefont {A.~P.}\ \bibnamefont
  {Bart\'ok}}, \ and\ \bibinfo {author} {\bibfnamefont {G.}~\bibnamefont
  {Cs\'anyi}},\ }\href {\doibase 10.1103/PhysRevB.90.104108} {\bibfield
  {journal} {\bibinfo  {journal} {Phys. Rev. B}\ }\textbf {\bibinfo {volume}
  {90}},\ \bibinfo {pages} {104108} (\bibinfo {year} {2014})}\BibitemShut
  {NoStop}%
\bibitem [{\citenamefont {Bart\'ok}\ \emph {et~al.}(2018)\citenamefont
  {Bart\'ok}, \citenamefont {Kermode}, \citenamefont {Bernstein},\ and\
  \citenamefont {Cs\'anyi}}]{PhysRevX.8.041048}%
  \BibitemOpen
  \bibfield  {author} {\bibinfo {author} {\bibfnamefont {A.~P.}\ \bibnamefont
  {Bart\'ok}}, \bibinfo {author} {\bibfnamefont {J.}~\bibnamefont {Kermode}},
  \bibinfo {author} {\bibfnamefont {N.}~\bibnamefont {Bernstein}}, \ and\
  \bibinfo {author} {\bibfnamefont {G.}~\bibnamefont {Cs\'anyi}},\ }\href
  {\doibase 10.1103/PhysRevX.8.041048} {\bibfield  {journal} {\bibinfo
  {journal} {Phys. Rev. X}\ }\textbf {\bibinfo {volume} {8}},\ \bibinfo {pages}
  {041048} (\bibinfo {year} {2018})}\BibitemShut {NoStop}%
\bibitem [{\citenamefont {Li}\ \emph {et~al.}(2015)\citenamefont {Li},
  \citenamefont {Kermode},\ and\ \citenamefont
  {De~Vita}}]{PhysRevLett.114.096405}%
  \BibitemOpen
  \bibfield  {author} {\bibinfo {author} {\bibfnamefont {Z.}~\bibnamefont
  {Li}}, \bibinfo {author} {\bibfnamefont {J.~R.}\ \bibnamefont {Kermode}}, \
  and\ \bibinfo {author} {\bibfnamefont {A.}~\bibnamefont {De~Vita}},\ }\href
  {\doibase 10.1103/PhysRevLett.114.096405} {\bibfield  {journal} {\bibinfo
  {journal} {Phys. Rev. Lett.}\ }\textbf {\bibinfo {volume} {114}},\ \bibinfo
  {pages} {096405} (\bibinfo {year} {2015})}\BibitemShut {NoStop}%
\bibitem [{\citenamefont {Glielmo}\ \emph {et~al.}(2017)\citenamefont
  {Glielmo}, \citenamefont {Sollich},\ and\ \citenamefont
  {De~Vita}}]{PhysRevB.95.214302}%
  \BibitemOpen
  \bibfield  {author} {\bibinfo {author} {\bibfnamefont {A.}~\bibnamefont
  {Glielmo}}, \bibinfo {author} {\bibfnamefont {P.}~\bibnamefont {Sollich}}, \
  and\ \bibinfo {author} {\bibfnamefont {A.}~\bibnamefont {De~Vita}},\ }\href
  {\doibase 10.1103/PhysRevB.95.214302} {\bibfield  {journal} {\bibinfo
  {journal} {Phys. Rev. B}\ }\textbf {\bibinfo {volume} {95}},\ \bibinfo
  {pages} {214302} (\bibinfo {year} {2017})}\BibitemShut {NoStop}%
\bibitem [{\citenamefont {Seko}\ \emph {et~al.}(2014)\citenamefont {Seko},
  \citenamefont {Takahashi},\ and\ \citenamefont
  {Tanaka}}]{PhysRevB.90.024101}%
  \BibitemOpen
  \bibfield  {author} {\bibinfo {author} {\bibfnamefont {A.}~\bibnamefont
  {Seko}}, \bibinfo {author} {\bibfnamefont {A.}~\bibnamefont {Takahashi}}, \
  and\ \bibinfo {author} {\bibfnamefont {I.}~\bibnamefont {Tanaka}},\ }\href
  {\doibase 10.1103/PhysRevB.90.024101} {\bibfield  {journal} {\bibinfo
  {journal} {Phys. Rev. B}\ }\textbf {\bibinfo {volume} {90}},\ \bibinfo
  {pages} {024101} (\bibinfo {year} {2014})}\BibitemShut {NoStop}%
\bibitem [{\citenamefont {Seko}\ \emph {et~al.}(2015)\citenamefont {Seko},
  \citenamefont {Takahashi},\ and\ \citenamefont
  {Tanaka}}]{PhysRevB.92.054113}%
  \BibitemOpen
  \bibfield  {author} {\bibinfo {author} {\bibfnamefont {A.}~\bibnamefont
  {Seko}}, \bibinfo {author} {\bibfnamefont {A.}~\bibnamefont {Takahashi}}, \
  and\ \bibinfo {author} {\bibfnamefont {I.}~\bibnamefont {Tanaka}},\ }\href
  {\doibase 10.1103/PhysRevB.92.054113} {\bibfield  {journal} {\bibinfo
  {journal} {Phys. Rev. B}\ }\textbf {\bibinfo {volume} {92}},\ \bibinfo
  {pages} {054113} (\bibinfo {year} {2015})}\BibitemShut {NoStop}%
\bibitem [{\citenamefont {Takahashi}\ \emph {et~al.}(2017)\citenamefont
  {Takahashi}, \citenamefont {Seko},\ and\ \citenamefont
  {Tanaka}}]{PhysRevMaterials.1.063801}%
  \BibitemOpen
  \bibfield  {author} {\bibinfo {author} {\bibfnamefont {A.}~\bibnamefont
  {Takahashi}}, \bibinfo {author} {\bibfnamefont {A.}~\bibnamefont {Seko}}, \
  and\ \bibinfo {author} {\bibfnamefont {I.}~\bibnamefont {Tanaka}},\ }\href
  {\doibase 10.1103/PhysRevMaterials.1.063801} {\bibfield  {journal} {\bibinfo
  {journal} {Phys. Rev. Mater.}\ }\textbf {\bibinfo {volume} {1}},\ \bibinfo
  {pages} {063801} (\bibinfo {year} {2017})}\BibitemShut {NoStop}%
\bibitem [{\citenamefont {Thompson}\ \emph {et~al.}(2015)\citenamefont
  {Thompson}, \citenamefont {Swiler}, \citenamefont {Trott}, \citenamefont
  {Foiles},\ and\ \citenamefont {Tucker}}]{Thompson2015316}%
  \BibitemOpen
  \bibfield  {author} {\bibinfo {author} {\bibfnamefont {A.}~\bibnamefont
  {Thompson}}, \bibinfo {author} {\bibfnamefont {L.}~\bibnamefont {Swiler}},
  \bibinfo {author} {\bibfnamefont {C.}~\bibnamefont {Trott}}, \bibinfo
  {author} {\bibfnamefont {S.}~\bibnamefont {Foiles}}, \ and\ \bibinfo {author}
  {\bibfnamefont {G.}~\bibnamefont {Tucker}},\ }\href {\doibase
  https://doi.org/10.1016/j.jcp.2014.12.018} {\bibfield  {journal} {\bibinfo
  {journal} {J. Comput. Phys.}\ }\textbf {\bibinfo {volume} {285}},\ \bibinfo
  {pages} {316 } (\bibinfo {year} {2015})}\BibitemShut {NoStop}%
\bibitem [{\citenamefont {Wood}\ and\ \citenamefont
  {Thompson}(2018)}]{wood2018extending}%
  \BibitemOpen
  \bibfield  {author} {\bibinfo {author} {\bibfnamefont {M.~A.}\ \bibnamefont
  {Wood}}\ and\ \bibinfo {author} {\bibfnamefont {A.~P.}\ \bibnamefont
  {Thompson}},\ }\href@noop {} {\bibfield  {journal} {\bibinfo  {journal} {J.
  Chem. Phys.}\ }\textbf {\bibinfo {volume} {148}},\ \bibinfo {pages} {241721}
  (\bibinfo {year} {2018})}\BibitemShut {NoStop}%
\bibitem [{\citenamefont {Chen}\ \emph {et~al.}(2017)\citenamefont {Chen},
  \citenamefont {Deng}, \citenamefont {Tran}, \citenamefont {Tang},
  \citenamefont {Chu},\ and\ \citenamefont {Ong}}]{PhysRevMaterials.1.043603}%
  \BibitemOpen
  \bibfield  {author} {\bibinfo {author} {\bibfnamefont {C.}~\bibnamefont
  {Chen}}, \bibinfo {author} {\bibfnamefont {Z.}~\bibnamefont {Deng}}, \bibinfo
  {author} {\bibfnamefont {R.}~\bibnamefont {Tran}}, \bibinfo {author}
  {\bibfnamefont {H.}~\bibnamefont {Tang}}, \bibinfo {author} {\bibfnamefont
  {I.-H.}\ \bibnamefont {Chu}}, \ and\ \bibinfo {author} {\bibfnamefont
  {S.~P.}\ \bibnamefont {Ong}},\ }\href {\doibase
  10.1103/PhysRevMaterials.1.043603} {\bibfield  {journal} {\bibinfo  {journal}
  {Phys. Rev. Mater.}\ }\textbf {\bibinfo {volume} {1}},\ \bibinfo {pages}
  {043603} (\bibinfo {year} {2017})}\BibitemShut {NoStop}%
\bibitem [{\citenamefont {Shapeev}(2016)}]{doi-10.1137-15M1054183}%
  \BibitemOpen
  \bibfield  {author} {\bibinfo {author} {\bibfnamefont {A.~V.}\ \bibnamefont
  {Shapeev}},\ }\href {\doibase 10.1137/15M1054183} {\bibfield  {journal}
  {\bibinfo  {journal} {Multiscale Model. Simul.}\ }\textbf {\bibinfo {volume}
  {14}},\ \bibinfo {pages} {1153} (\bibinfo {year} {2016})}\BibitemShut
  {NoStop}%
\bibitem [{\citenamefont {Mueller}\ \emph {et~al.}(2020)\citenamefont
  {Mueller}, \citenamefont {Hernandez},\ and\ \citenamefont
  {Wang}}]{doi:10.1063/1.5126336}%
  \BibitemOpen
  \bibfield  {author} {\bibinfo {author} {\bibfnamefont {T.}~\bibnamefont
  {Mueller}}, \bibinfo {author} {\bibfnamefont {A.}~\bibnamefont {Hernandez}},
  \ and\ \bibinfo {author} {\bibfnamefont {C.}~\bibnamefont {Wang}},\ }\href
  {\doibase 10.1063/1.5126336} {\bibfield  {journal} {\bibinfo  {journal} {J.
  Chem. Phys.}\ }\textbf {\bibinfo {volume} {152}},\ \bibinfo {pages} {050902}
  (\bibinfo {year} {2020})}\BibitemShut {NoStop}%
\bibitem [{\citenamefont {Khorshidi}\ and\ \citenamefont
  {Peterson}(2016)}]{khorshidi2016amp}%
  \BibitemOpen
  \bibfield  {author} {\bibinfo {author} {\bibfnamefont {A.}~\bibnamefont
  {Khorshidi}}\ and\ \bibinfo {author} {\bibfnamefont {A.~A.}\ \bibnamefont
  {Peterson}},\ }\href@noop {} {\bibfield  {journal} {\bibinfo  {journal}
  {Comput. Phys. Commun.}\ }\textbf {\bibinfo {volume} {207}},\ \bibinfo
  {pages} {310} (\bibinfo {year} {2016})}\BibitemShut {NoStop}%
\bibitem [{\citenamefont {Ferr\'e}\ \emph {et~al.}(2015)\citenamefont
  {Ferr\'e}, \citenamefont {Maillet},\ and\ \citenamefont
  {Stoltz}}]{doi-10.1063-1.4930541}%
  \BibitemOpen
  \bibfield  {author} {\bibinfo {author} {\bibfnamefont {G.}~\bibnamefont
  {Ferr\'e}}, \bibinfo {author} {\bibfnamefont {J.-B.}\ \bibnamefont
  {Maillet}}, \ and\ \bibinfo {author} {\bibfnamefont {G.}~\bibnamefont
  {Stoltz}},\ }\href {\doibase 10.1063/1.4930541} {\bibfield  {journal}
  {\bibinfo  {journal} {J. Chem. Phys.}\ }\textbf {\bibinfo {volume} {143}},\
  \bibinfo {pages} {104114} (\bibinfo {year} {2015})}\BibitemShut {NoStop}%
\bibitem [{\citenamefont {Ghasemi}\ \emph {et~al.}(2015)\citenamefont
  {Ghasemi}, \citenamefont {Hofstetter}, \citenamefont {Saha},\ and\
  \citenamefont {Goedecker}}]{PhysRevB.92.045131}%
  \BibitemOpen
  \bibfield  {author} {\bibinfo {author} {\bibfnamefont {S.~A.}\ \bibnamefont
  {Ghasemi}}, \bibinfo {author} {\bibfnamefont {A.}~\bibnamefont {Hofstetter}},
  \bibinfo {author} {\bibfnamefont {S.}~\bibnamefont {Saha}}, \ and\ \bibinfo
  {author} {\bibfnamefont {S.}~\bibnamefont {Goedecker}},\ }\href {\doibase
  10.1103/PhysRevB.92.045131} {\bibfield  {journal} {\bibinfo  {journal} {Phys.
  Rev. B}\ }\textbf {\bibinfo {volume} {92}},\ \bibinfo {pages} {045131}
  (\bibinfo {year} {2015})}\BibitemShut {NoStop}%
\bibitem [{\citenamefont {Botu}\ and\ \citenamefont
  {Ramprasad}(2015)}]{QUA:QUA24836}%
  \BibitemOpen
  \bibfield  {author} {\bibinfo {author} {\bibfnamefont {V.}~\bibnamefont
  {Botu}}\ and\ \bibinfo {author} {\bibfnamefont {R.}~\bibnamefont
  {Ramprasad}},\ }\href {\doibase 10.1002/qua.24836} {\bibfield  {journal}
  {\bibinfo  {journal} {Int. J. Quantum Chem.}\ }\textbf {\bibinfo {volume}
  {115}},\ \bibinfo {pages} {1074} (\bibinfo {year} {2015})}\BibitemShut
  {NoStop}%
\bibitem [{\citenamefont {Freitas}\ and\ \citenamefont
  {Cao}(2022)}]{Freitas2022}%
  \BibitemOpen
  \bibfield  {author} {\bibinfo {author} {\bibfnamefont {R.}~\bibnamefont
  {Freitas}}\ and\ \bibinfo {author} {\bibfnamefont {Y.}~\bibnamefont {Cao}},\
  }\href {\doibase 10.1557/s43579-022-00221-5} {\bibfield  {journal} {\bibinfo
  {journal} {MRS Commun.}\ } (\bibinfo {year} {2022}),\
  10.1557/s43579-022-00221-5}\BibitemShut {NoStop}%
\bibitem [{\citenamefont {Drautz}(2019)}]{PhysRevB.99.014104}%
  \BibitemOpen
  \bibfield  {author} {\bibinfo {author} {\bibfnamefont {R.}~\bibnamefont
  {Drautz}},\ }\href {\doibase 10.1103/PhysRevB.99.014104} {\bibfield
  {journal} {\bibinfo  {journal} {Phys. Rev. B}\ }\textbf {\bibinfo {volume}
  {99}},\ \bibinfo {pages} {014104} (\bibinfo {year} {2019})}\BibitemShut
  {NoStop}%
\bibitem [{\citenamefont {Deringer}\ \emph {et~al.}(2018)\citenamefont
  {Deringer}, \citenamefont {Pickard},\ and\ \citenamefont
  {Cs\'anyi}}]{PhysRevLett.120.156001}%
  \BibitemOpen
  \bibfield  {author} {\bibinfo {author} {\bibfnamefont {V.~L.}\ \bibnamefont
  {Deringer}}, \bibinfo {author} {\bibfnamefont {C.~J.}\ \bibnamefont
  {Pickard}}, \ and\ \bibinfo {author} {\bibfnamefont {G.}~\bibnamefont
  {Cs\'anyi}},\ }\href {\doibase 10.1103/PhysRevLett.120.156001} {\bibfield
  {journal} {\bibinfo  {journal} {Phys. Rev. Lett.}\ }\textbf {\bibinfo
  {volume} {120}},\ \bibinfo {pages} {156001} (\bibinfo {year}
  {2018})}\BibitemShut {NoStop}%
\bibitem [{\citenamefont {Podryabinkin}\ \emph {et~al.}(2019)\citenamefont
  {Podryabinkin}, \citenamefont {Tikhonov}, \citenamefont {Shapeev},\ and\
  \citenamefont {Oganov}}]{PhysRevB.99.064114}%
  \BibitemOpen
  \bibfield  {author} {\bibinfo {author} {\bibfnamefont {E.~V.}\ \bibnamefont
  {Podryabinkin}}, \bibinfo {author} {\bibfnamefont {E.~V.}\ \bibnamefont
  {Tikhonov}}, \bibinfo {author} {\bibfnamefont {A.~V.}\ \bibnamefont
  {Shapeev}}, \ and\ \bibinfo {author} {\bibfnamefont {A.~R.}\ \bibnamefont
  {Oganov}},\ }\href {\doibase 10.1103/PhysRevB.99.064114} {\bibfield
  {journal} {\bibinfo  {journal} {Phys. Rev. B}\ }\textbf {\bibinfo {volume}
  {99}},\ \bibinfo {pages} {064114} (\bibinfo {year} {2019})}\BibitemShut
  {NoStop}%
\bibitem [{\citenamefont {Gubaev}\ \emph {et~al.}(2019)\citenamefont {Gubaev},
  \citenamefont {Podryabinkin}, \citenamefont {Hart},\ and\ \citenamefont
  {Shapeev}}]{GUBAEV2019148}%
  \BibitemOpen
  \bibfield  {author} {\bibinfo {author} {\bibfnamefont {K.}~\bibnamefont
  {Gubaev}}, \bibinfo {author} {\bibfnamefont {E.~V.}\ \bibnamefont
  {Podryabinkin}}, \bibinfo {author} {\bibfnamefont {G.~L.}\ \bibnamefont
  {Hart}}, \ and\ \bibinfo {author} {\bibfnamefont {A.~V.}\ \bibnamefont
  {Shapeev}},\ }\href {\doibase
  https://doi.org/10.1016/j.commatsci.2018.09.031} {\bibfield  {journal}
  {\bibinfo  {journal} {Comput. Mater. Sci.}\ }\textbf {\bibinfo {volume}
  {156}},\ \bibinfo {pages} {148 } (\bibinfo {year} {2019})}\BibitemShut
  {NoStop}%
\bibitem [{\citenamefont {Kharabadze}\ \emph {et~al.}(2022)\citenamefont
  {Kharabadze}, \citenamefont {Thorn}, \citenamefont {Koulakova},\ and\
  \citenamefont {Kolmogorov}}]{Kharabadze2022}%
  \BibitemOpen
  \bibfield  {author} {\bibinfo {author} {\bibfnamefont {S.}~\bibnamefont
  {Kharabadze}}, \bibinfo {author} {\bibfnamefont {A.}~\bibnamefont {Thorn}},
  \bibinfo {author} {\bibfnamefont {E.~A.}\ \bibnamefont {Koulakova}}, \ and\
  \bibinfo {author} {\bibfnamefont {A.~N.}\ \bibnamefont {Kolmogorov}},\ }\href
  {\doibase 10.1038/s41524-022-00825-4} {\bibfield  {journal} {\bibinfo
  {journal} {npj Comput. Mater.}\ }\textbf {\bibinfo {volume} {8}},\ \bibinfo
  {pages} {136} (\bibinfo {year} {2022})}\BibitemShut {NoStop}%
\bibitem [{\citenamefont {Wakai}\ \emph {et~al.}(2023)\citenamefont {Wakai},
  \citenamefont {Seko},\ and\ \citenamefont {Tanaka}}]{wakai2023efficient}%
  \BibitemOpen
  \bibfield  {author} {\bibinfo {author} {\bibfnamefont {H.}~\bibnamefont
  {Wakai}}, \bibinfo {author} {\bibfnamefont {A.}~\bibnamefont {Seko}}, \ and\
  \bibinfo {author} {\bibfnamefont {I.}~\bibnamefont {Tanaka}},\ }\href
  {\doibase https://doi.org/10.2109/jcersj2.23053} {\bibfield  {journal}
  {\bibinfo  {journal} {J. Ceram. Soc. Japan}\ }\textbf {\bibinfo {volume}
  {131}},\ \bibinfo {pages} {762} (\bibinfo {year} {2023})}\BibitemShut
  {NoStop}%
\bibitem [{\citenamefont {Thorn}\ \emph {et~al.}(2023)\citenamefont {Thorn},
  \citenamefont {Gochitashvili}, \citenamefont {Kharabadze},\ and\
  \citenamefont {Kolmogorov}}]{D3CP02817H}%
  \BibitemOpen
  \bibfield  {author} {\bibinfo {author} {\bibfnamefont {A.}~\bibnamefont
  {Thorn}}, \bibinfo {author} {\bibfnamefont {D.}~\bibnamefont
  {Gochitashvili}}, \bibinfo {author} {\bibfnamefont {S.}~\bibnamefont
  {Kharabadze}}, \ and\ \bibinfo {author} {\bibfnamefont {A.~N.}\ \bibnamefont
  {Kolmogorov}},\ }\href {\doibase 10.1039/D3CP02817H} {\bibfield  {journal}
  {\bibinfo  {journal} {Phys. Chem. Chem. Phys.}\ }\textbf {\bibinfo {volume}
  {25}},\ \bibinfo {pages} {22415} (\bibinfo {year} {2023})}\BibitemShut
  {NoStop}%
\bibitem [{\citenamefont {Oganov}\ \emph {et~al.}(2019)\citenamefont {Oganov},
  \citenamefont {Pickard}, \citenamefont {Zhu},\ and\ \citenamefont
  {Needs}}]{oganov2019structure}%
  \BibitemOpen
  \bibfield  {author} {\bibinfo {author} {\bibfnamefont {A.~R.}\ \bibnamefont
  {Oganov}}, \bibinfo {author} {\bibfnamefont {C.~J.}\ \bibnamefont {Pickard}},
  \bibinfo {author} {\bibfnamefont {Q.}~\bibnamefont {Zhu}}, \ and\ \bibinfo
  {author} {\bibfnamefont {R.~J.}\ \bibnamefont {Needs}},\ }\href@noop {}
  {\bibfield  {journal} {\bibinfo  {journal} {Nature Rev. Mater.}\ }\textbf
  {\bibinfo {volume} {4}},\ \bibinfo {pages} {331} (\bibinfo {year}
  {2019})}\BibitemShut {NoStop}%
\bibitem [{\citenamefont {Pickard}\ and\ \citenamefont
  {Needs}(2011)}]{Pickard_2011}%
  \BibitemOpen
  \bibfield  {author} {\bibinfo {author} {\bibfnamefont {C.~J.}\ \bibnamefont
  {Pickard}}\ and\ \bibinfo {author} {\bibfnamefont {R.~J.}\ \bibnamefont
  {Needs}},\ }\href {\doibase 10.1088/0953-8984/23/5/053201} {\bibfield
  {journal} {\bibinfo  {journal} {J. Phys.: Condens. Mattter}\ }\textbf
  {\bibinfo {volume} {23}},\ \bibinfo {pages} {053201} (\bibinfo {year}
  {2011})}\BibitemShut {NoStop}%
\bibitem [{\citenamefont {Wang}\ \emph {et~al.}(2012)\citenamefont {Wang},
  \citenamefont {Lv}, \citenamefont {Zhu},\ and\ \citenamefont
  {Ma}}]{WANG20122063}%
  \BibitemOpen
  \bibfield  {author} {\bibinfo {author} {\bibfnamefont {Y.}~\bibnamefont
  {Wang}}, \bibinfo {author} {\bibfnamefont {J.}~\bibnamefont {Lv}}, \bibinfo
  {author} {\bibfnamefont {L.}~\bibnamefont {Zhu}}, \ and\ \bibinfo {author}
  {\bibfnamefont {Y.}~\bibnamefont {Ma}},\ }\href {\doibase
  https://doi.org/10.1016/j.cpc.2012.05.008} {\bibfield  {journal} {\bibinfo
  {journal} {Comput. Phys. Commun.}\ }\textbf {\bibinfo {volume} {183}},\
  \bibinfo {pages} {2063} (\bibinfo {year} {2012})}\BibitemShut {NoStop}%
\bibitem [{\citenamefont {Glass}\ \emph {et~al.}(2006)\citenamefont {Glass},
  \citenamefont {Oganov},\ and\ \citenamefont {Hansen}}]{GLASS2006713}%
  \BibitemOpen
  \bibfield  {author} {\bibinfo {author} {\bibfnamefont {C.~W.}\ \bibnamefont
  {Glass}}, \bibinfo {author} {\bibfnamefont {A.~R.}\ \bibnamefont {Oganov}}, \
  and\ \bibinfo {author} {\bibfnamefont {N.}~\bibnamefont {Hansen}},\ }\href
  {\doibase https://doi.org/10.1016/j.cpc.2006.07.020} {\bibfield  {journal}
  {\bibinfo  {journal} {Comput. Phys. Commun.}\ }\textbf {\bibinfo {volume}
  {175}},\ \bibinfo {pages} {713} (\bibinfo {year} {2006})}\BibitemShut
  {NoStop}%
\bibitem [{\citenamefont {Boender}\ and\ \citenamefont
  {Rinnooy~Kan}(1987)}]{boender1987bayesian}%
  \BibitemOpen
  \bibfield  {author} {\bibinfo {author} {\bibfnamefont {C.}~\bibnamefont
  {Boender}}\ and\ \bibinfo {author} {\bibfnamefont {A.}~\bibnamefont
  {Rinnooy~Kan}},\ }\href@noop {} {\bibfield  {journal} {\bibinfo  {journal}
  {Mathematical Programming}\ }\textbf {\bibinfo {volume} {37}},\ \bibinfo
  {pages} {59} (\bibinfo {year} {1987})}\BibitemShut {NoStop}%
\bibitem [{\citenamefont {Seko}\ \emph {et~al.}(2019)\citenamefont {Seko},
  \citenamefont {Togo},\ and\ \citenamefont {Tanaka}}]{PhysRevB.99.214108}%
  \BibitemOpen
  \bibfield  {author} {\bibinfo {author} {\bibfnamefont {A.}~\bibnamefont
  {Seko}}, \bibinfo {author} {\bibfnamefont {A.}~\bibnamefont {Togo}}, \ and\
  \bibinfo {author} {\bibfnamefont {I.}~\bibnamefont {Tanaka}},\ }\href
  {\doibase 10.1103/PhysRevB.99.214108} {\bibfield  {journal} {\bibinfo
  {journal} {Phys. Rev. B}\ }\textbf {\bibinfo {volume} {99}},\ \bibinfo
  {pages} {214108} (\bibinfo {year} {2019})}\BibitemShut {NoStop}%
\bibitem [{\citenamefont {Seko}(2020)}]{PhysRevB.102.174104}%
  \BibitemOpen
  \bibfield  {author} {\bibinfo {author} {\bibfnamefont {A.}~\bibnamefont
  {Seko}},\ }\href {\doibase 10.1103/PhysRevB.102.174104} {\bibfield  {journal}
  {\bibinfo  {journal} {Phys. Rev. B}\ }\textbf {\bibinfo {volume} {102}},\
  \bibinfo {pages} {174104} (\bibinfo {year} {2020})}\BibitemShut {NoStop}%
\bibitem [{\citenamefont {Seko}(2023)}]{doi:10.1063/5.0129045}%
  \BibitemOpen
  \bibfield  {author} {\bibinfo {author} {\bibfnamefont {A.}~\bibnamefont
  {Seko}},\ }\href {\doibase 10.1063/5.0129045} {\bibfield  {journal} {\bibinfo
   {journal} {J. Appl. Phys.}\ }\textbf {\bibinfo {volume} {133}},\ \bibinfo
  {pages} {011101} (\bibinfo {year} {2023})}\BibitemShut {NoStop}%
\bibitem [{\citenamefont {El-Batanouny}\ and\ \citenamefont
  {Wooten}(2008)}]{el-batanouny_wooten_2008}%
  \BibitemOpen
  \bibfield  {author} {\bibinfo {author} {\bibfnamefont {M.}~\bibnamefont
  {El-Batanouny}}\ and\ \bibinfo {author} {\bibfnamefont {F.}~\bibnamefont
  {Wooten}},\ }\href {\doibase 10.1017/CBO9780511755736} {\emph {\bibinfo
  {title} {Symmetry and Condensed Matter Physics: A Computational Approach}}}\
  (\bibinfo  {publisher} {Cambridge University Press},\ \bibinfo {year}
  {2008})\BibitemShut {NoStop}%
\bibitem [{\citenamefont {Bergerhoff}\ and\ \citenamefont
  {Brown}(1987)}]{bergerhoff1987crystal}%
  \BibitemOpen
  \bibfield  {author} {\bibinfo {author} {\bibfnamefont {G.}~\bibnamefont
  {Bergerhoff}}\ and\ \bibinfo {author} {\bibfnamefont {I.~D.}\ \bibnamefont
  {Brown}},\ }in\ \href@noop {} {\emph {\bibinfo {booktitle} {Crystallographic
  Databases}}},\ \bibinfo {editor} {edited by\ \bibinfo {editor} {\bibfnamefont
  {F.~H.}\ \bibnamefont {Allen~et al.}}}\ (\bibinfo  {publisher} {International
  Union of Crystallography, Chester},\ \bibinfo {year} {1987})\BibitemShut
  {NoStop}%
\bibitem [{\citenamefont {Bl{\"o}chl}(1994)}]{PAW1}%
  \BibitemOpen
  \bibfield  {author} {\bibinfo {author} {\bibfnamefont {P.~E.}\ \bibnamefont
  {Bl{\"o}chl}},\ }\href@noop {} {\bibfield  {journal} {\bibinfo  {journal}
  {Phys. Rev. B}\ }\textbf {\bibinfo {volume} {50}},\ \bibinfo {pages} {17953}
  (\bibinfo {year} {1994})}\BibitemShut {NoStop}%
\bibitem [{\citenamefont {Perdew}\ \emph {et~al.}(1996)\citenamefont {Perdew},
  \citenamefont {Burke},\ and\ \citenamefont {Ernzerhof}}]{GGA:PBE96}%
  \BibitemOpen
  \bibfield  {author} {\bibinfo {author} {\bibfnamefont {J.~P.}\ \bibnamefont
  {Perdew}}, \bibinfo {author} {\bibfnamefont {K.}~\bibnamefont {Burke}}, \
  and\ \bibinfo {author} {\bibfnamefont {M.}~\bibnamefont {Ernzerhof}},\
  }\href@noop {} {\bibfield  {journal} {\bibinfo  {journal} {Phys. Rev. Lett.}\
  }\textbf {\bibinfo {volume} {77}},\ \bibinfo {pages} {3865} (\bibinfo {year}
  {1996})}\BibitemShut {NoStop}%
\bibitem [{\citenamefont {Kresse}\ and\ \citenamefont {Hafner}(1993)}]{VASP1}%
  \BibitemOpen
  \bibfield  {author} {\bibinfo {author} {\bibfnamefont {G.}~\bibnamefont
  {Kresse}}\ and\ \bibinfo {author} {\bibfnamefont {J.}~\bibnamefont
  {Hafner}},\ }\href@noop {} {\bibfield  {journal} {\bibinfo  {journal} {Phys.
  Rev. B}\ }\textbf {\bibinfo {volume} {47}},\ \bibinfo {pages} {558} (\bibinfo
  {year} {1993})}\BibitemShut {NoStop}%
\bibitem [{\citenamefont {Kresse}\ and\ \citenamefont
  {Furthm{\"u}ller}(1996)}]{VASP2}%
  \BibitemOpen
  \bibfield  {author} {\bibinfo {author} {\bibfnamefont {G.}~\bibnamefont
  {Kresse}}\ and\ \bibinfo {author} {\bibfnamefont {J.}~\bibnamefont
  {Furthm{\"u}ller}},\ }\href@noop {} {\bibfield  {journal} {\bibinfo
  {journal} {Phys. Rev. B}\ }\textbf {\bibinfo {volume} {54}},\ \bibinfo
  {pages} {11169} (\bibinfo {year} {1996})}\BibitemShut {NoStop}%
\bibitem [{\citenamefont {Kresse}\ and\ \citenamefont {Joubert}(1999)}]{PAW2}%
  \BibitemOpen
  \bibfield  {author} {\bibinfo {author} {\bibfnamefont {G.}~\bibnamefont
  {Kresse}}\ and\ \bibinfo {author} {\bibfnamefont {D.}~\bibnamefont
  {Joubert}},\ }\href@noop {} {\bibfield  {journal} {\bibinfo  {journal} {Phys.
  Rev. B}\ }\textbf {\bibinfo {volume} {59}},\ \bibinfo {pages} {1758}
  (\bibinfo {year} {1999})}\BibitemShut {NoStop}%
\bibitem [{\citenamefont {Hendrix}\ \emph {et~al.}(2010)\citenamefont
  {Hendrix}, \citenamefont {Bogl{\'a}rka} \emph
  {et~al.}}]{hendrix2010introduction}%
  \BibitemOpen
  \bibfield  {author} {\bibinfo {author} {\bibfnamefont {E.~M.}\ \bibnamefont
  {Hendrix}}, \bibinfo {author} {\bibfnamefont {G.}~\bibnamefont
  {Bogl{\'a}rka}},  \emph {et~al.},\ }\href@noop {} {\emph {\bibinfo {title}
  {Introduction to nonlinear and global optimization}}},\ Vol.~\bibinfo
  {volume} {37}\ (\bibinfo  {publisher} {Springer},\ \bibinfo {year}
  {2010})\BibitemShut {NoStop}%
\bibitem [{\citenamefont {IUCr}(2002)}]{ITA2002}%
  \BibitemOpen
  \bibfield  {author} {\bibinfo {author} {\bibnamefont {IUCr}},\ }\href@noop {}
  {\emph {\bibinfo {title} {International Tables for Crystallography, Volume A:
  Space Group Symmetry}}},\ \bibinfo {edition} {5th}\ ed.,\ International
  Tables for Crystallography\ (\bibinfo  {publisher} {Kluwer Academic
  Publishers},\ \bibinfo {address} {Dordrecht, Boston, London},\ \bibinfo
  {year} {2002})\BibitemShut {NoStop}%
\bibitem [{\citenamefont {Valle}\ and\ \citenamefont
  {Oganov}(2010)}]{valle2010crystal}%
  \BibitemOpen
  \bibfield  {author} {\bibinfo {author} {\bibfnamefont {M.}~\bibnamefont
  {Valle}}\ and\ \bibinfo {author} {\bibfnamefont {A.~R.}\ \bibnamefont
  {Oganov}},\ }\href@noop {} {\bibfield  {journal} {\bibinfo  {journal} {Acta
  Crystallogr. A}\ }\textbf {\bibinfo {volume} {66}},\ \bibinfo {pages} {507}
  (\bibinfo {year} {2010})}\BibitemShut {NoStop}%
\bibitem [{Lam()}]{LammpsPolyMLP}%
  \BibitemOpen
  \href {https://github.com/sekocha/lammps-polymlp-package} {}\bibinfo {note}
  {{A. Seko}, {lammps-polymlp-package},
  \url{https://github.com/sekocha/lammps-polymlp-package}}\BibitemShut
  {NoStop}%
\bibitem [{Mac()}]{MachineLearningPotentialRepository}%
  \BibitemOpen
  \href {https://sekocha.github.io} {}\bibinfo {note} {{A. Seko}, {Polynomial}
  {Machine} {Learning} {Potential} {Repository} at {Kyoto} {University},
  \url{https://sekocha.github.io}}\BibitemShut {NoStop}%
\bibitem [{\citenamefont {Wakai}\ \emph {et~al.}(2024)\citenamefont {Wakai},
  \citenamefont {Seko}, \citenamefont {Izuta}, \citenamefont {Nishiyama},\ and\
  \citenamefont {Tanaka}}]{PhysRevB.109.214207}%
  \BibitemOpen
  \bibfield  {author} {\bibinfo {author} {\bibfnamefont {H.}~\bibnamefont
  {Wakai}}, \bibinfo {author} {\bibfnamefont {A.}~\bibnamefont {Seko}},
  \bibinfo {author} {\bibfnamefont {H.}~\bibnamefont {Izuta}}, \bibinfo
  {author} {\bibfnamefont {T.}~\bibnamefont {Nishiyama}}, \ and\ \bibinfo
  {author} {\bibfnamefont {I.}~\bibnamefont {Tanaka}},\ }\href {\doibase
  10.1103/PhysRevB.109.214207} {\bibfield  {journal} {\bibinfo  {journal}
  {Phys. Rev. B}\ }\textbf {\bibinfo {volume} {109}},\ \bibinfo {pages}
  {214207} (\bibinfo {year} {2024})}\BibitemShut {NoStop}%
\bibitem [{\citenamefont {Borg}\ and\ \citenamefont
  {Groenen}(2005)}]{borg2005modern}%
  \BibitemOpen
  \bibfield  {author} {\bibinfo {author} {\bibfnamefont {I.}~\bibnamefont
  {Borg}}\ and\ \bibinfo {author} {\bibfnamefont {P.~J.}\ \bibnamefont
  {Groenen}},\ }\href@noop {} {\emph {\bibinfo {title} {Modern multidimensional
  scaling: Theory and applications}}}\ (\bibinfo  {publisher} {Springer Science
  \& Business Media},\ \bibinfo {year} {2005})\BibitemShut {NoStop}%
\bibitem [{\citenamefont {Schiferl}\ and\ \citenamefont
  {Barrett}(1969)}]{schiferl1969crystal}%
  \BibitemOpen
  \bibfield  {author} {\bibinfo {author} {\bibfnamefont {D.}~\bibnamefont
  {Schiferl}}\ and\ \bibinfo {author} {\bibfnamefont {C.~S.}\ \bibnamefont
  {Barrett}},\ }\href@noop {} {\bibfield  {journal} {\bibinfo  {journal} {J.
  Appl. Crystallogr.}\ }\textbf {\bibinfo {volume} {2}},\ \bibinfo {pages} {30}
  (\bibinfo {year} {1969})}\BibitemShut {NoStop}%
\bibitem [{\citenamefont {Young}(1991)}]{young1991phase}%
  \BibitemOpen
  \bibfield  {author} {\bibinfo {author} {\bibfnamefont {D.~A.}\ \bibnamefont
  {Young}},\ }\href@noop {} {\emph {\bibinfo {title} {Phase diagrams of the
  elements}}}\ (\bibinfo  {publisher} {Univ of California Press},\ \bibinfo
  {year} {1991})\BibitemShut {NoStop}%
\bibitem [{\citenamefont {Smith}\ \emph {et~al.}(1975)\citenamefont {Smith},
  \citenamefont {Leadbetter},\ and\ \citenamefont
  {Apling}}]{smith1975structures}%
  \BibitemOpen
  \bibfield  {author} {\bibinfo {author} {\bibfnamefont {P.~M.}\ \bibnamefont
  {Smith}}, \bibinfo {author} {\bibfnamefont {A.~J.}\ \bibnamefont
  {Leadbetter}}, \ and\ \bibinfo {author} {\bibfnamefont {A.~J.}\ \bibnamefont
  {Apling}},\ }\href@noop {} {\bibfield  {journal} {\bibinfo  {journal} {Phil.
  Mag.}\ }\textbf {\bibinfo {volume} {31}},\ \bibinfo {pages} {57} (\bibinfo
  {year} {1975})}\BibitemShut {NoStop}%
\bibitem [{\citenamefont {Davey}(1925)}]{PhysRev.25.753}%
  \BibitemOpen
  \bibfield  {author} {\bibinfo {author} {\bibfnamefont {W.~P.}\ \bibnamefont
  {Davey}},\ }\href {\doibase 10.1103/PhysRev.25.753} {\bibfield  {journal}
  {\bibinfo  {journal} {Phys. Rev.}\ }\textbf {\bibinfo {volume} {25}},\
  \bibinfo {pages} {753} (\bibinfo {year} {1925})}\BibitemShut {NoStop}%
\bibitem [{\citenamefont {Cucka}\ and\ \citenamefont
  {Barrett}(1962)}]{cucka1962crystal}%
  \BibitemOpen
  \bibfield  {author} {\bibinfo {author} {\bibfnamefont {P.}~\bibnamefont
  {Cucka}}\ and\ \bibinfo {author} {\bibfnamefont {C.~S.}\ \bibnamefont
  {Barrett}},\ }\href@noop {} {\bibfield  {journal} {\bibinfo  {journal} {Acta
  Crystallogr.}\ }\textbf {\bibinfo {volume} {15}},\ \bibinfo {pages} {865}
  (\bibinfo {year} {1962})}\BibitemShut {NoStop}%
\bibitem [{\citenamefont {Jaggi}(1964)}]{jaggi1964struktur}%
  \BibitemOpen
  \bibfield  {author} {\bibinfo {author} {\bibfnamefont {R.}~\bibnamefont
  {Jaggi}},\ }in\ \href@noop {} {\emph {\bibinfo {booktitle} {Helvetica Physica
  Acta}}},\ Vol.~\bibinfo {volume} {37}\ (\bibinfo {year} {1964})\ p.\ \bibinfo
  {pages} {618}\BibitemShut {NoStop}%
\bibitem [{\citenamefont {Akselrud}\ \emph {et~al.}(2003)\citenamefont
  {Akselrud}, \citenamefont {Hanflandm},\ and\ \citenamefont
  {Schwarz}}]{AkselrudHanflandmSchwarz+2003+447+448}%
  \BibitemOpen
  \bibfield  {author} {\bibinfo {author} {\bibfnamefont {L.~G.}\ \bibnamefont
  {Akselrud}}, \bibinfo {author} {\bibfnamefont {M.}~\bibnamefont {Hanflandm}},
  \ and\ \bibinfo {author} {\bibfnamefont {U.}~\bibnamefont {Schwarz}},\ }\href
  {\doibase doi:10.1524/ncrs.2003.218.jg.447} {\bibfield  {journal} {\bibinfo
  {journal} {Z. Krist. - New Cryst. St.}\ }\textbf {\bibinfo {volume} {218}},\
  \bibinfo {pages} {447} (\bibinfo {year} {2003})}\BibitemShut {NoStop}%
\bibitem [{\citenamefont {Schaufelberger}\ \emph {et~al.}(1973)\citenamefont
  {Schaufelberger}, \citenamefont {Merx},\ and\ \citenamefont
  {Contre}}]{Schaufelberger1973}%
  \BibitemOpen
  \bibfield  {author} {\bibinfo {author} {\bibfnamefont {P.}~\bibnamefont
  {Schaufelberger}}, \bibinfo {author} {\bibfnamefont {H.}~\bibnamefont
  {Merx}}, \ and\ \bibinfo {author} {\bibfnamefont {M.}~\bibnamefont
  {Contre}},\ }\href@noop {} {\bibfield  {journal} {\bibinfo  {journal} {High
  Temp. High Press.}\ }\textbf {\bibinfo {volume} {5}},\ \bibinfo {pages} {221}
  (\bibinfo {year} {1973})}\BibitemShut {NoStop}%
\bibitem [{\citenamefont {Degtyareva}\ \emph {et~al.}(2001)\citenamefont
  {Degtyareva}, \citenamefont {McMahon},\ and\ \citenamefont
  {Nelmes}}]{degtyareva2001crystal}%
  \BibitemOpen
  \bibfield  {author} {\bibinfo {author} {\bibfnamefont {O.}~\bibnamefont
  {Degtyareva}}, \bibinfo {author} {\bibfnamefont {M.~I.}\ \bibnamefont
  {McMahon}}, \ and\ \bibinfo {author} {\bibfnamefont {R.~J.}\ \bibnamefont
  {Nelmes}},\ }in\ \href@noop {} {\emph {\bibinfo {booktitle} {Mater. Sci.
  Forum}}},\ Vol.\ \bibinfo {volume} {378}\ (\bibinfo {organization} {Trans
  Tech Publ},\ \bibinfo {year} {2001})\ pp.\ \bibinfo {pages}
  {469--475}\BibitemShut {NoStop}%
\bibitem [{\citenamefont {Chaimayo}\ \emph {et~al.}(2012)\citenamefont
  {Chaimayo}, \citenamefont {Lundegaard}, \citenamefont {Loa}, \citenamefont
  {Stinton}, \citenamefont {Lennie},\ and\ \citenamefont
  {McMahon}}]{chaimayo2012high}%
  \BibitemOpen
  \bibfield  {author} {\bibinfo {author} {\bibfnamefont {W.}~\bibnamefont
  {Chaimayo}}, \bibinfo {author} {\bibfnamefont {L.~F.}\ \bibnamefont
  {Lundegaard}}, \bibinfo {author} {\bibfnamefont {I.}~\bibnamefont {Loa}},
  \bibinfo {author} {\bibfnamefont {G.~W.}\ \bibnamefont {Stinton}}, \bibinfo
  {author} {\bibfnamefont {A.~R.}\ \bibnamefont {Lennie}}, \ and\ \bibinfo
  {author} {\bibfnamefont {M.~I.}\ \bibnamefont {McMahon}},\ }\href@noop {}
  {\bibfield  {journal} {\bibinfo  {journal} {High Press. Res.}\ }\textbf
  {\bibinfo {volume} {32}},\ \bibinfo {pages} {442} (\bibinfo {year}
  {2012})}\BibitemShut {NoStop}%
\bibitem [{\citenamefont {Kabalkina}\ \emph {et~al.}(1970)\citenamefont
  {Kabalkina}, \citenamefont {Kolobyanina},\ and\ \citenamefont
  {Vereshchagin}}]{kabalkina1970investigation}%
  \BibitemOpen
  \bibfield  {author} {\bibinfo {author} {\bibfnamefont {S.~S.}\ \bibnamefont
  {Kabalkina}}, \bibinfo {author} {\bibfnamefont {T.~N.}\ \bibnamefont
  {Kolobyanina}}, \ and\ \bibinfo {author} {\bibfnamefont {L.~F.}\ \bibnamefont
  {Vereshchagin}},\ }\href@noop {} {\bibfield  {journal} {\bibinfo  {journal}
  {Sov. Phys. JETP}\ }\textbf {\bibinfo {volume} {31}},\ \bibinfo {pages} {259}
  (\bibinfo {year} {1970})}\BibitemShut {NoStop}%
\bibitem [{\citenamefont {Sharma}\ and\ \citenamefont
  {Donohue}(1962)}]{+1962+293+300}%
  \BibitemOpen
  \bibfield  {author} {\bibinfo {author} {\bibfnamefont {B.~D.}\ \bibnamefont
  {Sharma}}\ and\ \bibinfo {author} {\bibfnamefont {J.}~\bibnamefont
  {Donohue}},\ }\href {\doibase doi:10.1524/zkri.1962.117.4.293} {\bibfield
  {journal} {\bibinfo  {journal} {Z. Krist.}\ }\textbf {\bibinfo {volume}
  {117}},\ \bibinfo {pages} {293} (\bibinfo {year} {1962})}\BibitemShut
  {NoStop}%
\bibitem [{\citenamefont {de~Koning}\ \emph {et~al.}(2009)\citenamefont
  {de~Koning}, \citenamefont {Antonelli},\ and\ \citenamefont
  {Jara}}]{PhysRevB.80.045209}%
  \BibitemOpen
  \bibfield  {author} {\bibinfo {author} {\bibfnamefont {M.}~\bibnamefont
  {de~Koning}}, \bibinfo {author} {\bibfnamefont {A.}~\bibnamefont
  {Antonelli}}, \ and\ \bibinfo {author} {\bibfnamefont {D.~A.~C.}\
  \bibnamefont {Jara}},\ }\href {\doibase 10.1103/PhysRevB.80.045209}
  {\bibfield  {journal} {\bibinfo  {journal} {Phys. Rev. B}\ }\textbf {\bibinfo
  {volume} {80}},\ \bibinfo {pages} {045209} (\bibinfo {year}
  {2009})}\BibitemShut {NoStop}%
\bibitem [{\citenamefont {Defrain}\ \emph {et~al.}(1961)\citenamefont
  {Defrain}, \citenamefont {Curien},\ and\ \citenamefont
  {Rimsky}}]{persee.fr:bulmi_0037-9328_1961_num_84_3_5486}%
  \BibitemOpen
  \bibfield  {author} {\bibinfo {author} {\bibfnamefont {A.}~\bibnamefont
  {Defrain}}, \bibinfo {author} {\bibfnamefont {H.}~\bibnamefont {Curien}}, \
  and\ \bibinfo {author} {\bibfnamefont {A.}~\bibnamefont {Rimsky}},\ }\href
  {\doibase 10.3406/bulmi.1961.5486} {\bibfield  {journal} {\bibinfo  {journal}
  {Bull. Minéral.}\ }\textbf {\bibinfo {volume} {84}},\ \bibinfo {pages} {260}
  (\bibinfo {year} {1961})}\BibitemShut {NoStop}%
\bibitem [{\citenamefont {Bosio}(2008)}]{10.1063/1.435841}%
  \BibitemOpen
  \bibfield  {author} {\bibinfo {author} {\bibfnamefont {L.}~\bibnamefont
  {Bosio}},\ }\href {\doibase 10.1063/1.435841} {\bibfield  {journal} {\bibinfo
   {journal} {J. Chem. Phys.}\ }\textbf {\bibinfo {volume} {68}},\ \bibinfo
  {pages} {1221} (\bibinfo {year} {2008})}\BibitemShut {NoStop}%
\bibitem [{\citenamefont {Bosio}\ \emph {et~al.}()\citenamefont {Bosio},
  \citenamefont {Curien}, \citenamefont {Dupont},\ and\ \citenamefont
  {Rimsky}}]{https://doi.org/10.1107/S0567740873002530}%
  \BibitemOpen
  \bibfield  {author} {\bibinfo {author} {\bibfnamefont {L.}~\bibnamefont
  {Bosio}}, \bibinfo {author} {\bibfnamefont {H.}~\bibnamefont {Curien}},
  \bibinfo {author} {\bibfnamefont {M.}~\bibnamefont {Dupont}}, \ and\ \bibinfo
  {author} {\bibfnamefont {A.}~\bibnamefont {Rimsky}},\ }\href {\doibase
  https://doi.org/10.1107/S0567740873002530} {\bibfield  {journal} {\bibinfo
  {journal} {Acta Crystallogr. B}\ }\textbf {\bibinfo {volume} {29}},\ \bibinfo
  {pages} {367}}\BibitemShut {NoStop}%
\bibitem [{\citenamefont {Hull}\ and\ \citenamefont
  {Davey}(1921)}]{PhysRev.17.549}%
  \BibitemOpen
  \bibfield  {author} {\bibinfo {author} {\bibfnamefont {A.~W.}\ \bibnamefont
  {Hull}}\ and\ \bibinfo {author} {\bibfnamefont {W.~P.}\ \bibnamefont
  {Davey}},\ }\href {\doibase 10.1103/PhysRev.17.549} {\bibfield  {journal}
  {\bibinfo  {journal} {Phys. Rev.}\ }\textbf {\bibinfo {volume} {17}},\
  \bibinfo {pages} {549} (\bibinfo {year} {1921})}\BibitemShut {NoStop}%
\bibitem [{\citenamefont {Takemura}\ and\ \citenamefont
  {Fujihisa}(1993)}]{PhysRevB.47.8465}%
  \BibitemOpen
  \bibfield  {author} {\bibinfo {author} {\bibfnamefont {K.}~\bibnamefont
  {Takemura}}\ and\ \bibinfo {author} {\bibfnamefont {H.}~\bibnamefont
  {Fujihisa}},\ }\href {\doibase 10.1103/PhysRevB.47.8465} {\bibfield
  {journal} {\bibinfo  {journal} {Phys. Rev. B}\ }\textbf {\bibinfo {volume}
  {47}},\ \bibinfo {pages} {8465} (\bibinfo {year} {1993})}\BibitemShut
  {NoStop}%
\bibitem [{\citenamefont {Spedding}\ \emph {et~al.}(1961)\citenamefont
  {Spedding}, \citenamefont {Hanak},\ and\ \citenamefont
  {Daane}}]{spedding1961high}%
  \BibitemOpen
  \bibfield  {author} {\bibinfo {author} {\bibfnamefont {F.~H.}\ \bibnamefont
  {Spedding}}, \bibinfo {author} {\bibfnamefont {J.~J.}\ \bibnamefont {Hanak}},
  \ and\ \bibinfo {author} {\bibfnamefont {A.~H.}\ \bibnamefont {Daane}},\
  }\href@noop {} {\bibfield  {journal} {\bibinfo  {journal} {J. Less-common
  Met.}\ }\textbf {\bibinfo {volume} {3}},\ \bibinfo {pages} {110} (\bibinfo
  {year} {1961})}\BibitemShut {NoStop}%
\bibitem [{\citenamefont {Nomura}\ \emph {et~al.}(1977)\citenamefont {Nomura},
  \citenamefont {Hayakawa},\ and\ \citenamefont {Ono}}]{nomura1977lanthanum}%
  \BibitemOpen
  \bibfield  {author} {\bibinfo {author} {\bibfnamefont {K.}~\bibnamefont
  {Nomura}}, \bibinfo {author} {\bibfnamefont {H.}~\bibnamefont {Hayakawa}}, \
  and\ \bibinfo {author} {\bibfnamefont {S.}~\bibnamefont {Ono}},\ }\href@noop
  {} {\bibfield  {journal} {\bibinfo  {journal} {J. Less-common Met.}\ }\textbf
  {\bibinfo {volume} {52}},\ \bibinfo {pages} {259} (\bibinfo {year}
  {1977})}\BibitemShut {NoStop}%
\bibitem [{\citenamefont {Brown}\ and\ \citenamefont
  {Rundqvist}(1965)}]{brown1965refinement}%
  \BibitemOpen
  \bibfield  {author} {\bibinfo {author} {\bibfnamefont {A.}~\bibnamefont
  {Brown}}\ and\ \bibinfo {author} {\bibfnamefont {S.}~\bibnamefont
  {Rundqvist}},\ }\href@noop {} {\bibfield  {journal} {\bibinfo  {journal}
  {Acta Crystallogr.}\ }\textbf {\bibinfo {volume} {19}},\ \bibinfo {pages}
  {684} (\bibinfo {year} {1965})}\BibitemShut {NoStop}%
\bibitem [{\citenamefont {Jamieson}(1963)}]{jamieson1963crystal}%
  \BibitemOpen
  \bibfield  {author} {\bibinfo {author} {\bibfnamefont {J.~C.}\ \bibnamefont
  {Jamieson}},\ }\href@noop {} {\bibfield  {journal} {\bibinfo  {journal}
  {Science}\ }\textbf {\bibinfo {volume} {139}},\ \bibinfo {pages} {1291}
  (\bibinfo {year} {1963})}\BibitemShut {NoStop}%
\bibitem [{\citenamefont {Kikegawa}\ and\ \citenamefont
  {Iwasaki}(1983)}]{kikegawa1983x}%
  \BibitemOpen
  \bibfield  {author} {\bibinfo {author} {\bibfnamefont {T.}~\bibnamefont
  {Kikegawa}}\ and\ \bibinfo {author} {\bibfnamefont {H.}~\bibnamefont
  {Iwasaki}},\ }\href@noop {} {\bibfield  {journal} {\bibinfo  {journal} {Acta
  Crystallogr. B}\ }\textbf {\bibinfo {volume} {39}},\ \bibinfo {pages} {158}
  (\bibinfo {year} {1983})}\BibitemShut {NoStop}%
\bibitem [{\citenamefont {Marqu\'es}\ \emph {et~al.}(2008)\citenamefont
  {Marqu\'es}, \citenamefont {Ackland}, \citenamefont {Lundegaard},
  \citenamefont {Falconi}, \citenamefont {Hejny}, \citenamefont {McMahon},
  \citenamefont {Contreras-Garc\'{\i}a},\ and\ \citenamefont
  {Hanfland}}]{PhysRevB.78.054120}%
  \BibitemOpen
  \bibfield  {author} {\bibinfo {author} {\bibfnamefont {M.}~\bibnamefont
  {Marqu\'es}}, \bibinfo {author} {\bibfnamefont {G.~J.}\ \bibnamefont
  {Ackland}}, \bibinfo {author} {\bibfnamefont {L.~F.}\ \bibnamefont
  {Lundegaard}}, \bibinfo {author} {\bibfnamefont {S.}~\bibnamefont {Falconi}},
  \bibinfo {author} {\bibfnamefont {C.}~\bibnamefont {Hejny}}, \bibinfo
  {author} {\bibfnamefont {M.~I.}\ \bibnamefont {McMahon}}, \bibinfo {author}
  {\bibfnamefont {J.}~\bibnamefont {Contreras-Garc\'{\i}a}}, \ and\ \bibinfo
  {author} {\bibfnamefont {M.}~\bibnamefont {Hanfland}},\ }\href {\doibase
  10.1103/PhysRevB.78.054120} {\bibfield  {journal} {\bibinfo  {journal} {Phys.
  Rev. B}\ }\textbf {\bibinfo {volume} {78}},\ \bibinfo {pages} {054120}
  (\bibinfo {year} {2008})}\BibitemShut {NoStop}%
\bibitem [{\citenamefont {Barrett}\ \emph {et~al.}(1963)\citenamefont
  {Barrett}, \citenamefont {Cucka},\ and\ \citenamefont
  {Haefner}}]{barrett1963crystal}%
  \BibitemOpen
  \bibfield  {author} {\bibinfo {author} {\bibfnamefont {C.~S.}\ \bibnamefont
  {Barrett}}, \bibinfo {author} {\bibfnamefont {P.}~\bibnamefont {Cucka}}, \
  and\ \bibinfo {author} {\bibfnamefont {K.~J. A.~C.}\ \bibnamefont
  {Haefner}},\ }\href@noop {} {\bibfield  {journal} {\bibinfo  {journal} {Acta
  Crystallogr.}\ }\textbf {\bibinfo {volume} {16}},\ \bibinfo {pages} {451}
  (\bibinfo {year} {1963})}\BibitemShut {NoStop}%
\bibitem [{\citenamefont {Vereschchagin}\ and\ \citenamefont
  {Kabalkina}(1965)}]{vereschchagin1965phase}%
  \BibitemOpen
  \bibfield  {author} {\bibinfo {author} {\bibfnamefont {L.~F.}\ \bibnamefont
  {Vereschchagin}}\ and\ \bibinfo {author} {\bibfnamefont {S.~S.}\ \bibnamefont
  {Kabalkina}},\ }\href@noop {} {\bibfield  {journal} {\bibinfo  {journal}
  {Sov. Phys. JETP}\ }\textbf {\bibinfo {volume} {20}},\ \bibinfo {pages} {274}
  (\bibinfo {year} {1965})}\BibitemShut {NoStop}%
\bibitem [{\citenamefont {Aoki}\ \emph {et~al.}(1983)\citenamefont {Aoki},
  \citenamefont {Fujiwara},\ and\ \citenamefont {Kusakabe}}]{aoki1983new}%
  \BibitemOpen
  \bibfield  {author} {\bibinfo {author} {\bibfnamefont {K.}~\bibnamefont
  {Aoki}}, \bibinfo {author} {\bibfnamefont {S.}~\bibnamefont {Fujiwara}}, \
  and\ \bibinfo {author} {\bibfnamefont {M.}~\bibnamefont {Kusakabe}},\
  }\href@noop {} {\bibfield  {journal} {\bibinfo  {journal} {Solid State
  Commun.}\ }\textbf {\bibinfo {volume} {45}},\ \bibinfo {pages} {161}
  (\bibinfo {year} {1983})}\BibitemShut {NoStop}%
\bibitem [{\citenamefont {Schwarz}\ \emph {et~al.}(2003)\citenamefont
  {Schwarz}, \citenamefont {Akselrud}, \citenamefont {Rosner}, \citenamefont
  {Ormeci}, \citenamefont {Grin},\ and\ \citenamefont
  {Hanfland}}]{PhysRevB.67.214101}%
  \BibitemOpen
  \bibfield  {author} {\bibinfo {author} {\bibfnamefont {U.}~\bibnamefont
  {Schwarz}}, \bibinfo {author} {\bibfnamefont {L.}~\bibnamefont {Akselrud}},
  \bibinfo {author} {\bibfnamefont {H.}~\bibnamefont {Rosner}}, \bibinfo
  {author} {\bibfnamefont {A.}~\bibnamefont {Ormeci}}, \bibinfo {author}
  {\bibfnamefont {Y.}~\bibnamefont {Grin}}, \ and\ \bibinfo {author}
  {\bibfnamefont {M.}~\bibnamefont {Hanfland}},\ }\href {\doibase
  10.1103/PhysRevB.67.214101} {\bibfield  {journal} {\bibinfo  {journal} {Phys.
  Rev. B}\ }\textbf {\bibinfo {volume} {67}},\ \bibinfo {pages} {214101}
  (\bibinfo {year} {2003})}\BibitemShut {NoStop}%
\bibitem [{\citenamefont {Swanson}(1953)}]{swanson1953standard}%
  \BibitemOpen
  \bibfield  {author} {\bibinfo {author} {\bibfnamefont {H.~E.}\ \bibnamefont
  {Swanson}},\ }\href@noop {} {\emph {\bibinfo {title} {Standard X-ray
  diffraction powder patterns}}},\ Vol.~\bibinfo {volume} {25}\ (\bibinfo
  {publisher} {US Department of Commerce, National Bureau of Standards},\
  \bibinfo {year} {1953})\BibitemShut {NoStop}%
\bibitem [{\citenamefont {Lee}\ and\ \citenamefont
  {Raynor}(1954)}]{lee1954lattice}%
  \BibitemOpen
  \bibfield  {author} {\bibinfo {author} {\bibfnamefont {J.~A.}\ \bibnamefont
  {Lee}}\ and\ \bibinfo {author} {\bibfnamefont {G.~V.}\ \bibnamefont
  {Raynor}},\ }\href@noop {} {\bibfield  {journal} {\bibinfo  {journal} {Proc.
  Phys. Soc. B}\ }\textbf {\bibinfo {volume} {67}},\ \bibinfo {pages} {737}
  (\bibinfo {year} {1954})}\BibitemShut {NoStop}%
\bibitem [{\citenamefont {Barnett}\ \emph {et~al.}(1966)\citenamefont
  {Barnett}, \citenamefont {Bean},\ and\ \citenamefont {Hall}}]{barnett1966x}%
  \BibitemOpen
  \bibfield  {author} {\bibinfo {author} {\bibfnamefont {J.~D.}\ \bibnamefont
  {Barnett}}, \bibinfo {author} {\bibfnamefont {V.~E.}\ \bibnamefont {Bean}}, \
  and\ \bibinfo {author} {\bibfnamefont {H.~T.}\ \bibnamefont {Hall}},\
  }\href@noop {} {\bibfield  {journal} {\bibinfo  {journal} {J. Appl. Phys.}\
  }\textbf {\bibinfo {volume} {37}},\ \bibinfo {pages} {875} (\bibinfo {year}
  {1966})}\BibitemShut {NoStop}%
\bibitem [{\citenamefont {Salamat}\ \emph {et~al.}(2013)\citenamefont
  {Salamat}, \citenamefont {Briggs}, \citenamefont {Bouvier}, \citenamefont
  {Petitgirard}, \citenamefont {Dewaele}, \citenamefont {Cutler}, \citenamefont
  {Cor\`a}, \citenamefont {Daisenberger}, \citenamefont {Garbarino},\ and\
  \citenamefont {McMillan}}]{PhysRevB.88.104104}%
  \BibitemOpen
  \bibfield  {author} {\bibinfo {author} {\bibfnamefont {A.}~\bibnamefont
  {Salamat}}, \bibinfo {author} {\bibfnamefont {R.}~\bibnamefont {Briggs}},
  \bibinfo {author} {\bibfnamefont {P.}~\bibnamefont {Bouvier}}, \bibinfo
  {author} {\bibfnamefont {S.}~\bibnamefont {Petitgirard}}, \bibinfo {author}
  {\bibfnamefont {A.}~\bibnamefont {Dewaele}}, \bibinfo {author} {\bibfnamefont
  {M.~E.}\ \bibnamefont {Cutler}}, \bibinfo {author} {\bibfnamefont
  {F.}~\bibnamefont {Cor\`a}}, \bibinfo {author} {\bibfnamefont
  {D.}~\bibnamefont {Daisenberger}}, \bibinfo {author} {\bibfnamefont
  {G.}~\bibnamefont {Garbarino}}, \ and\ \bibinfo {author} {\bibfnamefont
  {P.~F.}\ \bibnamefont {McMillan}},\ }\href {\doibase
  10.1103/PhysRevB.88.104104} {\bibfield  {journal} {\bibinfo  {journal} {Phys.
  Rev. B}\ }\textbf {\bibinfo {volume} {88}},\ \bibinfo {pages} {104104}
  (\bibinfo {year} {2013})}\BibitemShut {NoStop}%
\bibitem [{\citenamefont {Bradley}(1924)}]{doi:10.1080/14786442408634511}%
  \BibitemOpen
  \bibfield  {author} {\bibinfo {author} {\bibfnamefont {A.~J.}\ \bibnamefont
  {Bradley}},\ }\href {\doibase 10.1080/14786442408634511} {\bibfield
  {journal} {\bibinfo  {journal} {Lond. Edinb. Dublin Philos. Mag. J. Sci}\
  }\textbf {\bibinfo {volume} {48}},\ \bibinfo {pages} {477} (\bibinfo {year}
  {1924})}\BibitemShut {NoStop}%
\bibitem [{\citenamefont {Aoki}\ \emph {et~al.}(1980)\citenamefont {Aoki},
  \citenamefont {Shimomura},\ and\ \citenamefont {Minomura}}]{aoki1980crystal}%
  \BibitemOpen
  \bibfield  {author} {\bibinfo {author} {\bibfnamefont {K.}~\bibnamefont
  {Aoki}}, \bibinfo {author} {\bibfnamefont {O.}~\bibnamefont {Shimomura}}, \
  and\ \bibinfo {author} {\bibfnamefont {S.}~\bibnamefont {Minomura}},\
  }\href@noop {} {\bibfield  {journal} {\bibinfo  {journal} {J. Phys. Soc.
  Japan}\ }\textbf {\bibinfo {volume} {48}},\ \bibinfo {pages} {551} (\bibinfo
  {year} {1980})}\BibitemShut {NoStop}%
\bibitem [{\citenamefont {Takumi}\ \emph {et~al.}(2002)\citenamefont {Takumi},
  \citenamefont {Masamitsu},\ and\ \citenamefont {Nagata}}]{takumi2002x}%
  \BibitemOpen
  \bibfield  {author} {\bibinfo {author} {\bibfnamefont {M.}~\bibnamefont
  {Takumi}}, \bibinfo {author} {\bibfnamefont {T.}~\bibnamefont {Masamitsu}}, \
  and\ \bibinfo {author} {\bibfnamefont {K.}~\bibnamefont {Nagata}},\
  }\href@noop {} {\bibfield  {journal} {\bibinfo  {journal} {J. Phys.: Condens.
  Matter}\ }\textbf {\bibinfo {volume} {14}},\ \bibinfo {pages} {10609}
  (\bibinfo {year} {2002})}\BibitemShut {NoStop}%
\bibitem [{\citenamefont {Vezzoli}(1971)}]{vezzoli1971proposed}%
  \BibitemOpen
  \bibfield  {author} {\bibinfo {author} {\bibfnamefont {G.~C.}\ \bibnamefont
  {Vezzoli}},\ }\href@noop {} {\bibfield  {journal} {\bibinfo  {journal} {Z.
  Krist. - Cryst. Mater.}\ }\textbf {\bibinfo {volume} {134}},\ \bibinfo
  {pages} {305} (\bibinfo {year} {1971})}\BibitemShut {NoStop}%
\end{thebibliography}%

\end{document}